\documentclass{aa}
\newcommand\floor[1]{\lfloor#1\rfloor}
\usepackage{natbib}
\bibpunct{(}{)}{;}{a}{}{,} 
\usepackage{graphicx}
\usepackage{float}
\usepackage{subcaption}
\usepackage[normalem]{ulem}
\usepackage{txfonts}
\usepackage{hyperref}
\begin{document}   

\title{TFAW: wavelet-based signal reconstruction to reduce photometric noise in time-domain surveys}

   \author{D. del Ser
          \inst{1,2,3}
          \and
          O. Fors\inst{1,2,3}
          \and 
          J. N{\'u}\~{n}ez \inst{1,3}
          }

   \institute{Dept. de Física Quàntica i Astrofísica, Institut de Ciències del Cosmos (ICCUB),
Universitat de Barcelona, IEEC-UB, Mart\'{\i} i Franqu\`es 1, E08028 Barcelona, Spain\\
		\email{dser@fqa.ub.edu}
         \and
         University of North Carolina at Chapel Hill, Department of Physics and Astronomy, Chapel Hill, NC 27599-3255, USA
         \and
         Observatori Fabra, Reial Acadèmia de Ciències i Arts de Barcelona, Rambla dels Estudis, 115, E-08002 Barcelona, Spain
             }

   \titlerunning{Modification of TFA using wavelets}
   \authorrunning{D. del Ser el al.}
   
   \date{Received XXXX, XXXX; accepted XXXX, XXXX}

 
  \abstract
   {There have been many efforts to correct systematic effects in astronomical light curves to improve the detection and characterization of planetary transits and astrophysical variability. Algorithms like the Trend Filtering Algorithm (TFA) use simultaneously-observed stars to measure and remove systematic effects, and binning is used to reduce high-frequency random noise.}
   {We present TFAW, a wavelet-based modified version of TFA. TFAW aims first, to increase the periodic signal detection and second, to return a detrended and de-noised signal without modifying its intrinsic characteristics.}
   {We modify TFA's frequency analysis step adding a Stationary Wavelet Transform filter to perform an initial noise and outlier removal and increase the detection of variable signals. A wavelet-based filter is added to TFA's signal reconstruction to perform an adaptive characterization of the noise- and trend-free signal and the underlying noise contribution at each iteration while preserving astrophysical signals. We carried out tests over simulated sinusoidal and transit-like signals to assess the effectiveness of the method and applied TFAW to real light curves from TFRM. We also studied TFAW's application to simulated multiperiodic signals.}
   {TFAW improves the signal detection rate by increasing the signal detection efficiency (SDE) up to a factor $\sim$2.5$\times$ for low SNR light curves. For simulated transits, the transit detection rate improves by a factor $\sim$2-5$\times$ in the low-SNR regime compared to TFA. TFAW signal approximation performs up to a factor $\sim$2$\times$ better than bin averaging for planetary transits. The standard deviations of simulated and real TFAW light curves are $\sim$40$\%$ better compared to TFA. TFAW yields better MCMC posterior distributions and returns lower uncertainties, less biased transit parameters and narrower ( $\sim$10$\times$) credibility intervals for simulated transits. TFAW is also able to improve the characterization of multiperiodic signals. We present a newly-discovered variable star from TFRM.}
   {}

   
   \maketitle

\section{Introduction}

The photometric precision and accuracy achieved by an astronomical survey is a key factor in transit signal detection. The photometric precision that can be attained depends on several factors such as the atmosphere conditions, the instrument, the number of non-distorted stellar PSFs that can be imaged on the CCD, the reduction pipeline and the photometric errors of the measurements. Detectors with wide field of view (FoV) often suffer from additional issues such as undersampled PSFs, atmospheric extinction, source blending and optical distortion at the edges of the detector. These and other effects increase the noise of the light curves and also introduce systematic variations that decrease the detection probability of any periodic signal in the data.

Many of the systematic variations in a given light curve are shared by light curves of other stars in the same data set. In order to remove those systematics, one can identify the objects in the field that suffer from the same kind of variations as the target (correlated noise) and then build and apply a filter based on the light curves of these comparison stars. Trend Filtering Algorithm (hereafter TFA)~\citep{Kovacs2005} and SysRem \citep{Tamuz2005} are often applied to remove systematic variations in time-domain surveys, in particular wide FoV and/or multi-telescope transiting exoplanetary surveys, such as SuperWASP~\citep{Pollacco2006}, MEarth~\citep{irwin2009}, HAT-South~\citep{Bakos2009}, TFRM-PSES~\citep{Fors2013} or NGTS~\citep{Wheatley2013}. 

Wavelets have been applied in several astronomy areas like astronomical signal processing~\citep{Starck1994, Nunez1996, Starck1998}; redshift spectra study~\citep{Machado2013}; cosmic microwave background~\citep{Moudden2005}, baryon acoustic oscillation analysis~\citep{Arnalte2012}; solar activity~\citep{Aschwanden1998,Gimenez2001}; stellar activity, pulsation and rotation analysis~\citep{Freitas2010,Bravo2014}; signal detection~\citep{Szatmary1994,Fors2008,Otazu2002,Regulo2007}; galaxy distribution morphology~\citep{Antoja2012} and light curve noise analysis~\citep{Cubillos2017}, and filtering~\citep{Carter2009,Grziwa2014,Grziwa2016,Grziwa2016a,Waldmann2014}.

TFA implements correlated noise removal using multiple stars in the field. The first part of the algorithm consists of removing trends from trend- and noise-dominated time series in order to increase the probability of detecting weak signals (called \textit{frequency analysis} in the original article). Secondly, if a significant periodic signal is present in the target light curve, TFA reconstructs the shape of that signal using an iterative process (called \textit{signal reconstruction}). One key aspect of the signal reconstruction step is that it should minimize the chances of suppressing the signal or modifying its intrinsic astrophysical characteristics. Several efforts have extended TFA to improve the signal recovery, such as using a meticulous selection of the template stars (for example,~\citet{Kim2009} through hierarchical methods or Principal Component Analysis,~\citep{Petigura2012}, or taking into account stellar variability or multiperiodic signals~\citep{Kovacs2008}). TFA is further described in Sect~\ref{sect:tfa}.

The TFA modification we present here (hereafter TFAW) introduces a wavelet filter in order to 1) remove outliers using a wavelet-inferred estimation of the signal, 2) search for periods in this outlier-free and de-noised signal during the frequency analysis step, 3) estimate the shape of the trend- and noise-free phase folded signal and 4) iteratively de-noise the trend-free light curve during the signal reconstruction process. TFAW combines TFA detrending and systematics removal capabilities with the wavelet transform's signal decoupling and de-noising potential. The filter is built using the Stationary Wavelet Transform algorithm (hereafter SWT) or \textit{à trous} algorithm~\citep{Holschneider1989}. TFAW differs from other wavelet-based noise-filtering algorithms in that it does not require any parametric model fitting as in~\citet{Carter2009} nor any extra computational method~\citep{Waldmann2014}. Also, the noise contribution of the signal is estimated directly from its SWT at each iteration step and the de-noising is done through the subtraction of this contribution from the signal. This allows TFAW to de-noise the signal without modifying any of its intrinsic properties in contrast to wavelet coefficient thresholding~\citep{Grziwa2014,Grziwa2016} that can lead to distortions of the signal and introduce artificial oscillations (ripples) around discontinuities~\citep{Mallat2008}. The aim of this article is to compare the signal detection efficiency, the de-noising and signal reconstruction capabilities of TFAW and TFA, by assessing its performance on a set of simulated and real light curves. Both planet detection and characterization conducted in current and near-future surveys, such as CoRoT~\citep{Auvergne2009} and Kepler~\citep{Borucki2003}, could be direct candidates to benefit from TFAW improvements in signal detection and signal reconstruction without distortion shown in this work.

In Sects~\ref{sect:dwd} and \ref{subsect:wps} the SWT and the Wavelet Power Spectrum are defined. In Sects ~\ref{sect:tfaw} and~\ref{subsect:outliers_removal}, the TFAW and a SWT-based outliers removal algorithm are described, respectively. In Sects ~\ref{subseq:snlevel} and ~\ref{sect:motherwave} the selection criteria for the signal and noise levels and the mother wavelet are introduced.

In Sect~\ref{sect:tfawperf}, we apply TFA and TFAW to sinusoidal and transit-like simulated light curves affected by combinations of Gaussian and correlated noises. To assess the performance of both algorithms we compare them in a number of tests. First, in Sect~\ref{sect:transitdetect}, the TFA and TFAW planetary transit detection efficiency is evaluated. Second, the intrinsic signal shape and depth invariance after TFAW detrending and de-noising is validated in Sect~\ref{sect:signalrec}. Next, the application of the algorithm to multiperiodic superimposed signals is studied in Sect.~\ref{sect:multi}. In Sect.~\ref{sect:binvswave}, the better signal approximation of TFAW with respect of TFA bin averaging is shown. The performance of TFA and TFAW is assessed in terms of the bias of the fitted transit parameters values and their uncertainties in Sect.~\ref{sect:MCMC}.

We apply TFAW to light curves from a wide FoV ground-based data set: the Telescope Fabra-ROA at Montsec (TFRM) in Sect~\ref{sect:tfrm}, to confirm that TFAW outperforms TFA. 
One of the light curves detrended with TFAW from TFRM is found to be newly-discovered as variable star. In Sect.~\ref{sect:tfaw_quant}, a TFAW vs. TFA quantitative assessment is performed for a set of real light curves from TFRM.

\section{The TFAW algorithm}
\subsection{TFA basics and formulation}
\label{sect:tfa}

The key point behind TFA is the assumption that many stars within a photometric data set suffer from the same or similar systematic effects. Taking a representative set of template stars, one can build a filter function that the removal of correlated systematics from a target light curve. Given a target N-point light curve and a set of M zero-averaged template stars \{$X_j(i)$;$i=1,2,...,N$;$j=1,2,...,M$\}, the filter for that target is defined as:

   \begin{equation}\label{eq:1}
      F(i) = \sum\limits_{j=1}^M c_jX_j(i) ,
   \end{equation}

The set of coefficients \{$c_j$\} is determined by minimizing the following expression:

   \begin{equation}\label{eq:2}
      D = \sum\limits_{i=1}^N [Y(i) - A(i) - F(i)]^2 ,
   \end{equation}

Where $\{Y(i)\}$ represents the target light curve and $\{A(i)\}$ is either constant in the case of the frequency analysis step of the algorithm or, the best representation of the trend- and noise-free signal to be found in the iterative signal reconstruction step. Once the filter has been computed, the corrected light curve is defined by:

   \begin{equation}\label{eq:3}
      \hat{Y}(i) = Y(i) - \sum\limits_{j=1}^M c_jX_j(i)  ,
   \end{equation}
   
Although TFA is able to remove trends and systematics from photometric time series, this algorithm does not completely decouple stochastic and correlated noises from the underlying signals.

\subsection{The Discrete Wavelet Transform and the Stationary Wavelet Transform}
\label{sect:dwd}

Unlike the Fourier transform, the wavelet transform decomposes a given time series into its \textit{wavelets}, i.e., highly localized impulses obtained from scaling and shifting the \textit{mother wavelet} function. These scaling and shifting operations allow us to calculate the wavelet coefficients, which represent the correlation between the wavelet and a localized section of the signal. The wavelet coefficients are calculated for each wavelet segment, giving a time-series function measuring the wavelets' correlation to the signal. In comparison to the sine wave used in the Fourier transform, which is smooth and of infinite length, the wavelet is irregular in shape and compactly supported. These properties make wavelets an ideal tool for analyzing signals of a non-stationary nature. Their irregular shape allows us to analyze signals with discontinuities or sharp changes, while their compactly supported nature allows temporal localization of the signal's features.

In the Discrete Wavelet Transform (DWT) the signal is decomposed into dyadic blocks (i.e. shifting and scaling is based on a power 2). The DWT of a given signal is calculated by applying a series of filters to the signal. In the first level of decomposition the signal passes through a low pass filter resulting in a convolution of the original signal and the filter. Simultaneously, the signal is also decomposed using a high-pass filter. The outputs give us the \textit{detail coefficients} (related to the high-pass filter) and the \textit{approximation coefficients} (related to the low-pass filter). The filter outputs are then sub-sampled by 2. In an iterative process, the original signal is then successively decomposed into components of lower scale, while the high frequency components are retained but not further decomposed. The decomposition is represented as a binary tree with the nodes representing a sub-space with a different time-frequency localization. The tree is known as a filter bank while the wavelet multi-scale decomposition and reconstruction is known as multi-resolution analysis (MRA)~\citep{Mallat1989}. The maximum number of decomposition levels that can be performed is dependent on the input size of the data: a signal of length $2^k$ can be decomposed into k discrete levels or scales. 

The DWT is not a time-invariant transform and is therefore very sensitive to the alignment of the signal in time (the DWT of a translated version of a signal might not be the translated version of the DWT of the signal). The Stationary Wavelet Transform (SWT) is a wavelet transform algorithm designed to overcome the lack of translation-invariance of the DWT. As with the DWT, high and low pass filters are applied to the signal at each level. However, in order to achieve translation-invariance, the downsamplers and upsamplers are removed and the filter coefficients are upsampled by a factor of $2^{(k-1)}$ in the kth level of the algorithm. By padding the filters at each level with zeroes, the two new sequences at each level have the same length as the original time series. 

	\begin{figure}[h]
    \centering
    \includegraphics[width=\linewidth,height=50.0cm, keepaspectratio]{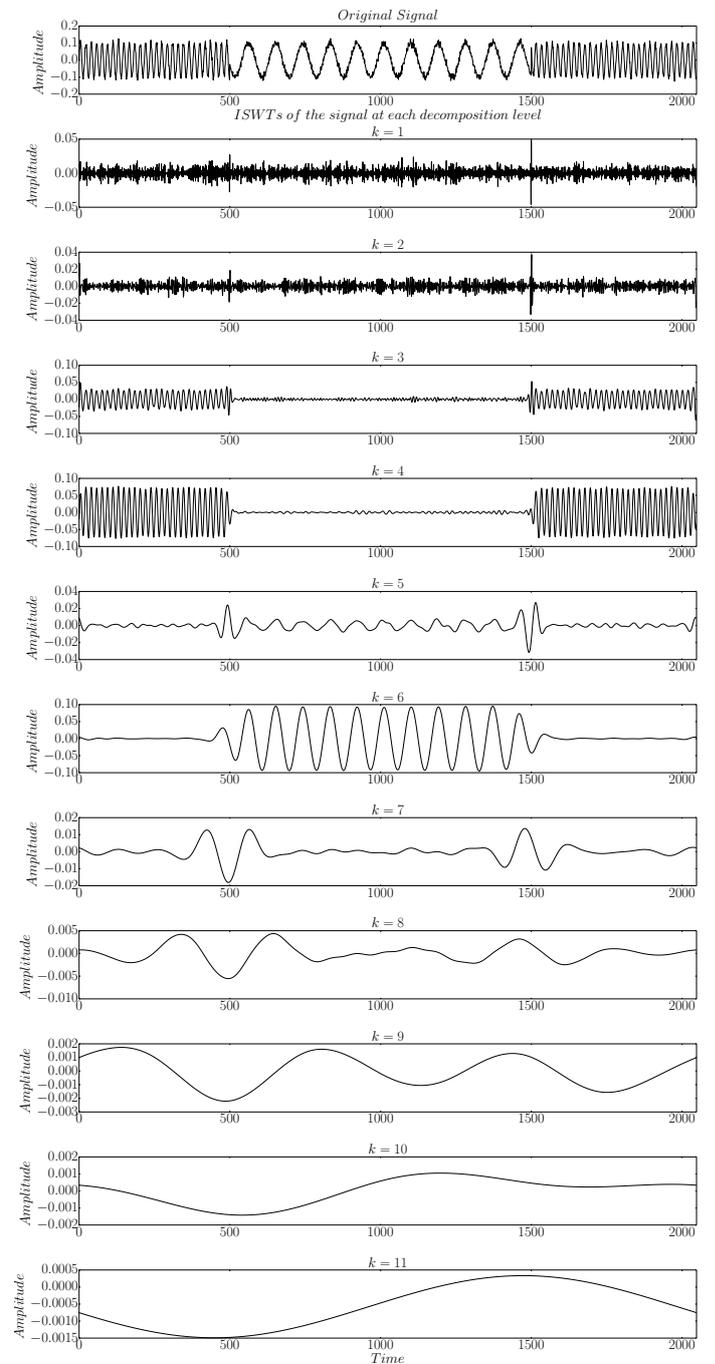}
    \caption{SWT decomposition of a test signal. The uppermost plot shows the test signal as described in Section \ref{sect:dwd}, consisting of two sinusoidal signals and Gaussian random noise. The following plots represent the ISWT transform of the signal at each SWT decomposition level (k = 1,...,11).}
	\label{SWTex}
	\end{figure}

An example of a MRA using the SWT is shown in Fig.~\ref{SWTex}. Our test signal is a combination of two sinusoidal signals, one of high frequency and another of low frequency. They both have the same amplitude and are affected by Gaussian random noise. As can be seen in Fig.~\ref{SWTex}, SWT is able to decouple each of these signal contributions at different decomposition levels. This is shown by computing the inverse SWT (hereafter ISWT) at each level. High-frequency noise has been separated from the sinusoidal signals, appearing as an extra contribution to the lower decomposition levels. In this example, noise can be fairly represented by the combination of the first two decomposition levels. The high frequency sinusoidal signal can be recovered by summing the third and fourth decomposition levels while the low frequency one would be recovered through the combination of the remaining levels. Crucially, the wavelet decomposition allows the signal characteristics to change through the time series, while retrieving both time and frequency information from the original signal. Other techniques such as the discrete Fourier Transform convert data from time into frequency domain but in doing so, time information is lost.

\subsection{Wavelet Power Spectrum}\label{sec:WavePowerSpec}
\label{subsect:wps}

The wavelet transform converts signal $f$ into a series of wavelet coefficients $W_{k}(t)$. Each of these coefficients represents the amplitude of the wavelet function at a particular location within the signal at a particular wavelet scale.
A useful way to determine the distribution of energy within the signal is to compute the wavelet power spectrum (WPS). The local WPS at a particular decomposition level $k$ is calculated by summing up the squares of wavelet coefficients for that level \citep{Torrence1998}:

\begin{equation}\label{eq:wpowerspect}
P(k) = \sum_{t} \mid W_{k}(t) \mid^{2}
\end{equation}

In the case of the Continuous Wavelet Transform (CWT), one can easily compute a relationship between the scale and the corresponding period. Following the method of~\citet{Meyers1993}, the relationship between the equivalent Fourier period and the wavelet scale can be derived analytically for a particular wavelet function, $\psi$, by substituting a cosine wave of a known frequency into the convolution of the mother wavelet with the Discrete Fourier Transform (DFT)~\citep{Torrence1998}. The use of the CWT power spectrum as a way to detect periodicities and study other astrophysical properties such as pulsation and rotation has already been studied by~\citet{Freitas2010} and~\citet{Bravo2014}.

When using the SWT to compute the WPS, the relationship between the scale and the frequency cannot be so easily computed as the signal energy can be distributed among different scales. However, the fact that the SWT power spectrum also presents peaks, will be of use for the outlier removal and signal detection methods described in Sect.~\ref{subsect:outliers_removal} and Sect.~\ref{subseq:snlevel}.

\subsection{TFAW: application of the SWT}
\label{sect:tfaw}

In the original version of TFA, once a significant periodic signal has been found in the target time series during the frequency analysis step, the phase-folded light curve is used to iteratively estimate $\{A(i)\}$ (the noise-free signal). They employ the simple bin average method, using a fixed number of bins (100 for $\sim$3000 data points) to ensure statistical stability and to obtain a reasonable noise averaging. They point out, however, that more accurate methods can be used to approximate $\{A(i)\}$. For example, in the case of planetary transits, one could use a~\citet{Mandel2002} model fit to get a more precise estimation of the noise- and trend-free signal.

In TFAW we follow a similar approach to the original TFA. First, an initial filter is computed using $\{A(i)\}$ = $\langle{Y}\rangle$. Then, the SWT and WPS of the filtered data $\hat{Y}(i)$ = $Y(i) - \{A(i)\}$ are computed. Using the method explained in Sect. \ref{subsect:outliers_removal} outliers are removed and an \textit{estimated signal} is obtained. The latter is then used to run Box Least Square (BLS)~\citep{Kovacs2002} (or Lomb-Scargle (LS)~\citet{Scargle1982}) to search for any periodic signal in our target light curve.

If any significant period is found, the light curve is phase folded to that period. Instead of using a bin average, $\{A(i)\}$ is estimated using the SWT of the phase folded signal. We decompose the phase folded light curve up to a SWT decomposition level given by the following equation:

   \begin{equation}\label{eq:4}
      max\ scale = \floor{\frac{\log(\frac{n}{l-1})}{\log 2}},
   \end{equation}

where n is the signal length and $l$ is the wavelet filter length.

At each decomposition level, we set all the \textit{detail coefficients} and \textit{approximation coefficients} to zero except the ones corresponding to that level. Then, the ISWT is computed for each level separately (as seen in Fig.~\ref{SWTex}). In order to estimate the shape of the noise- and trend-free signal we set a threshold decomposition level or \textit{signal level} (see details in Section \ref{subseq:snlevel}). $\{A(i)\}$ is then computed as the sum of the ISWTs of those levels from \textit{signal level} to the last decomposition level. In this way we ensure that our signal's estimate is separated from the high frequency noise while preserving the high-frequency shape of the signal (unlike normal binning; see Sect. \ref{sect:binvswave} for details in SWT signal approximation versus bin averaging): the high frequency noise components are better characterized by the lower decomposition levels as they have higher frequency resolution.

This new signal estimate, $\{A(i)\}$, can then be used to compute a new set of \{$c_j$\} coefficients by means of equation \ref{eq:2}. The new filter obtained from this set of coefficients gives us the new corrected light curve at each step of iteration through equation \ref{eq:3}. We set an additional filter at each iteration given by the SWT of the phase folded light curve. In this case, instead of using those decomposition levels above \textit{signal level} to reconstruct the noiseless signal, we set a \textit{noise level} so that the noise is characterized by the sum of the ISWTs of those levels below it. Thus equation \ref{eq:3} at each iteration step becomes:

   \begin{equation}\label{eq:5}
      \hat{Y'}(i) = \hat{Y}(i) - ISWT(\hat{Y(i)}, \textit{noise level}),
   \end{equation}

The iteration of the algorithm stops when the relative difference between the standard deviations of the residuals in the successive iterations falls under a certain limit. As with the original TFA algorithm, we set the limit at $10^{-3}$.

To summarize, the main steps of TFAW are the following:
\begin{enumerate}
\item An initial filter is computed using $\{A(i)\}$ = $\langle{Y}\rangle$ as with the original TFA to remove trends and other systematics.
\item The SWT and WPS are computed from the filtered light curve obtained in 1.
\item Outliers are removed and an outlier-free and de-noised \textit{estimated signal} is obtained.
\item The \textit{estimated signal} is used to search for periodicities.
\item If a significant period is found, the signal is phased folded to run the iterative signal reconstruction. Otherwise, no reconstruction is performed.
\item Using the SWT a new signal estimation $\{A(i)\}$ is computed by means of the \textit{signal level}. The noise contribution of the light curve is estimated using the \textit{noise level}.
\item The new $\{A(i)\}$ and noise contribution are used to compute the new filtered signal $\hat{Y'}(i)$.
\item Iteration continues until the convergence criterion is met.
\end{enumerate}

\subsection{Outlier Removal Using the SWT}
\label{subsect:outliers_removal}

An outlier can be defined as an observation point that is distant from the norm of the sample. Their origin can be diverse, but the presence of outliers in a data sample leads to effects in the mean which can display a bias toward the outlier value. Also, depending on their nature, they may impact the time series analysis with respect to modeling, estimation or forecasting. There are several methods for outlier detection: model-based ~\citep{Peirce1852,Grubbs1950}; interquartile range~\citep{Tukey1949}; distance-based~\citep{Knorr2000} and wavelet-based~\citep{Mallat1992,Bilen2002,Grane2010}.

One way of detecting outliers is through wavelet thresholding \citep{Bilen2002}. First, the wavelet transform of the target signal is computed. Then, a threshold value is computed at each decomposition level. Any wavelet coefficient that exceeds the threshold corresponding to that scale can be considered a potential outlier. Once the outliers have been detected, their associated coefficients are set to zero and the inverse wavelet transform is computed. However, this approach can introduce artificial oscillations (ripples) around outliers and discontinuities~\citep{Mallat2008,Grziwa2014,Grziwa2016} and also, if the threshold is not correctly computed, modifications to the signal's shape can result.

The outlier detection and removal approach developed for this work is a distance-based method that relies on an \textit{estimated signal}. 
As stated in Sect. \ref{sect:tfaw}, a noise-free signal can be obtained using the sum of the ISWT for those levels from the \textit{signal level} to the last decomposition level. Instead of using a pre-set \textit{signal level} to obtain our \textit{estimated signal}, we can define it using the WPS defined in Section \ref{subsect:wps}. We know that the WPS determines the distribution of energy within the signal. In the case of the SWT, it allows to determine the decomposition levels at which the correlation between the signal and the wavelet transform at that scale was higher. Thus, the scale of the highest peak in the WPS defines our \textit{signal level}. However, given that the nature of our light curves could be multi-periodic resulting in peaks at different scales, the peak at the lowest scale (i.e. highest frequency) is used to define the \textit{signal level}. In addition, given the discrete nature of the SWT and, also depending on the number of decomposition levels, the energy of our signal can be distributed within more than one scale. Thus, the \textit{signal level} is always set to also consider the previous scale in order to avoid losing as less signal contributions as possible. With this \textit{signal level} we can then reconstruct the \textit{estimated signal}, as explained before, by summing the corresponding ISWTs. A threshold can then be built in such a way that, given a point in the time series, if its distance to the \textit{estimated signal} is above the threshold, then, it is considered an outlier and removed (or replaced by any given value). In this work, the threshold value is computed using the \textit{Universal Threshold} \citep{Donoho1994b} and, every point exceeding 5 times this value, are considered as outliers. Once the  target light curve is free of outliers, the \textit{estimated signal} is recomputed. Except in the case of a very high frequency signal (for which the WPS peak at the lowest scale could coincide with the scale characterizing the noise), this final \textit{estimated signal} represents an outlier-free and de-noised approximation to our target signal that, as stated before, will be used to search for any periodicities within our light curve.
\begin{figure}[bth]
    \centering
    \includegraphics[width=\linewidth,height=10.0cm, keepaspectratio]{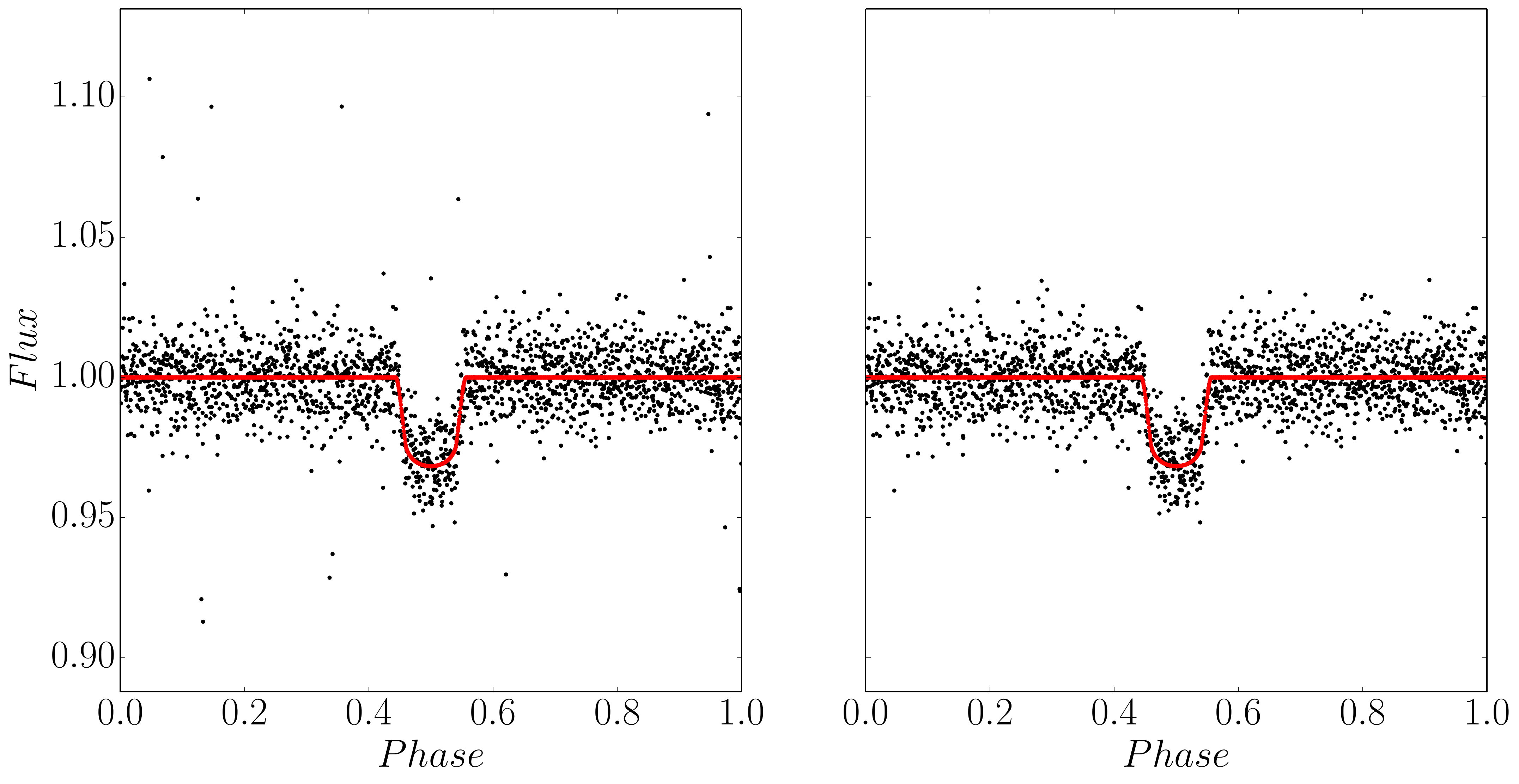}
    \caption{Outlier removal example. \textbf{Left:} Simulated phased folded light curve of Planet 1 in Table \ref{TransitParams} affected by outliers. Red line corresponds to the injected signal. \textbf{Right:} Same light curve with outliers removed.}
    \label{fig:OutlierRemoval}
\end{figure}

Fig.~\ref{fig:OutlierRemoval} shows an example of the outlier removal capabilities of the method. Left plot corresponds to a simulated phase folded light curve of Planet 1 in Table \ref{TransitParams}. Using the signal estimation explained above, once the threshold value has been estimated, the algorithm is able to effectively remove any outlier as seen in Fig.~\ref{fig:OutlierRemoval} right.

\subsection{Signal and noise level choice criteria}
\label{subseq:snlevel}

The selection of the \textit{signal level} and \textit{noise level} parameters must be carefully performed. As described above, high-frequency noise appears as an extra contribution to the lower decomposition levels of the SWT. However, high frequency signals hidden within the data (i.e. second order sinusoidal modulations, etc...) could be present in some of the levels in which the noise also appears. One way of solving this issue would be to increase the number of data points to a higher power of two. The greater the number of levels we decompose our signal, the easier it is to separate the noise contribution from the high frequency signals. Nonetheless, even if part of the signal is mixed with the noise at some level, using TFAW signal reconstruction capabilities, one could be able to recover the signal with almost any distortion given that the correct period has been found except in those cases in which the signal has exactly the same (or very close) frequency as the noise.

One way to diminish the signal contribution lost at each iteration step and avoid as much signal distortion is to select a \textit{noise level} value that uses just the lowest SWT decomposition levels (usually the first one), and accept the noise contribution from the lower decomposition levels. We recommend setting a \textit{noise level} equal to 1 (lowest SWT decomposition level) for time series shorter than 2048 points (i.e. 10 decomposition levels) and increase it with care for each extra power of two of the time series length.

Regarding the \textit{signal level} for the phase folded light curve, given that it should provide the best signal approximation, we recommend using the method explained in Sect. \ref{subsect:outliers_removal}. First computing the WPS of the phase folded signal (which should be already outlier-free) and then estimating the \textit{signal level} from the WPS' peaks.

\subsection{Mother wavelet selection criteria}
\label{sect:motherwave}
There are several mother wavelets with different analytic properties that can be used for signal decomposition. For our application of the SWT-modified version of TFA we use the non-orthogonal, symmetric base of the biorthogonal (bior) wavelet family, first proposed by~\citet{Cohen1992}. More specifically, we used the pre-computed values of the bior 3.9 wavelet included in the {\tt PyWavelets}\footnote{\url{https://pywavelets.readthedocs.io}} module. We selected this wavelet family for TFAW as it allows the construction of symmetrical wavelet functions and because the shape of the reconstruction scaling function is very similar to the characteristic shape of a planetary transit (searches for other types of astrophysical variability may benefit from other wavelet shapes).

\section{TFAW performance}
\label{sect:tfawperf}

In~\citet{Kovacs2005} several tests are presented demonstrating the signal detection and reconstruction capabilities of TFA. In this section we explore how the inclusion of the SWT filter in TFAW can improve the detection of variable signals, assess the impact that this filter has on the frequency power spectra, study the effects on the SNR of the light curves. Also we compare the use of the SWT to compute the signal approximation against bin averaging, test the likeness of the final TFAW light curve to the inserted variable signal and compare the results with the ones obtained using the original TFA. For all the simulated light curves hereafter and the real ones in Section \ref{sect:tfrm} we have used 2048 data points and set the \textit{noise} and \textit{signal levels} to the lower decomposition level and the one given by the WPS peak as explained in Sect. \ref{subsect:outliers_removal} and \ref{subseq:snlevel}.

\subsection{TFAW vs TFA transit detection efficiency}
\label{sect:transitdetect}

In order to assess the transit detection capabilities of BLS during TFAW and TFA frequency analysis step, we generate a set of 5000 time series with different random combinations of Gaussian and correlated noises. We also introduce linear and exponential trends (i.e. to simulate changes in airmass, changes in the CCD position of the object, etc), as well as gaps and jumps in the data. We use a set of 250 simulated template stars suffering from the same jumps, gaps and trends as the target time series, affected by random distributions of $\sigma_\mathrm{w}$ and $\sigma_\mathrm{r}$ combinations. To each of these light curves, we add the transit signal given by the parameters of planets 1 and 2 in Table~\ref{TransitParams} separately. We use {\tt batman}\footnote{\url{https://www.cfa.harvard.edu/~lkreidberg/batman}}~\cite{Kreidberg2015} package to simulate a~\citet{Mandel2002} planetary transit model. The total noise contribution for each light curve ranges from 0.005 to 0.2 mag for planet 1 and from 0.005 to 0.1 mag for planet 2.

   \begin{table}[h]
   \caption{Planet parameters used for TFAW simulations}             
   \label{TransitParams}      
   \centering                         
   \begin{tabular}{c c c c c c}       
   \hline\hline
   \noalign{\smallskip}
   Planet & $R_{P}\ (R_{J})$ & $M_{P}\ (M_{J})$ & $M_{*} \ (M_{\odot})$ & $R_{*}\ (R_{\odot})$ & P\ (d)\\     
   \hline                       
   \noalign{\smallskip}
   1 & 1.98 & 1.40 & 1.36 & 1.23 & 0.8468 \\     
   2 & 1.23 & 1.10 & 1.36 & 1.23 & 0.4842 \\
   \hline                                   
   \end{tabular}
   \end{table}

We want to compare the detection rates obtained with TFA and TFAW light curves. We use the same two criteria as~\citet{Kovacs2005} and define a detection as the following:
\begin{itemize}
\item The highest peak in the BLS power spectrum must have a frequency between $[\mathrm{f}_\mathrm{p}-0.001, \mathrm{f}_\mathrm{p}+0.001]$, where $\mathrm{f}_\mathrm{p}$ is the frequency of the simulated planetary transit.
\item The signal detection efficiency (SDE)~\citep{Kovacs2002} of the highest peak in the power spectrum must be greater than 6.
\end{itemize}
We run BLS using 100 bins and 99000 frequency steps to ensure statistical stability ~\citep{Kovacs2002}. The transit search is done in the (0.01, 12) days range.
We count the number of times the signal is detected in the TFAW light curves but not detected in the TFA ones. By performing this and the opposite test using a template sample of 250 stars, we obtain the results shown in Table~\ref{tfawdetect}.

   \begin{table}[h]
   \caption{Mutually exclusive detections and mean SDE values for simulated planetary transits 1 and 2 as shown in Table~\ref{TransitParams}. $\mathrm{N}_\mathrm{TFA}$: not detected using TFAW light curves, but detected using TFA data. $\mathrm{N}_\mathrm{TFAW}$: detected using TFAW light curves, but not detected using TFA data. $\mathrm{N}_\mathrm{mut}$: simultaneous detections with TFA and TFAW. $\mathrm{SDE}_\mathrm{TFA}$: mean TFA SDE. $\mathrm{SDE}_\mathrm{TFAW}$: mean TFAW SDE.  Percentage values in parenthesis are with respect to the 5000 tested transits.}
   \label{tfawdetect}      
   \centering
   \begin{tabular}{c c c c c c}       
   \hline\hline
   \noalign{\smallskip}
   Planet & $\mathrm{N}_\mathrm{TFA}$ & $\mathrm{N}_\mathrm{TFAW}$ & $\mathrm{N}_\mathrm{mut}$ & $\mathrm{SDE}_\mathrm{TFA}$ & $\mathrm{SDE}_\mathrm{TFAW}$\\     
   \hline                       
   \noalign{\smallskip}
   1 & 265 & 784 & 561 & 16.99 & 17.86 \\
     & (5.3$\%$) & (15.7$\%$) & (11.2$\%$) & & \\
   2 & 79 & 503 & 612 & 14.61 & 15.92 \\
     & (1.58$\%$) & (10.1$\%$) & (12.2$\%$) & & \\
   \hline                                   
   \end{tabular}
   \end{table}

\begin{figure}[bth]
    \centering
    \includegraphics[width=\linewidth,height=10.0cm, keepaspectratio]{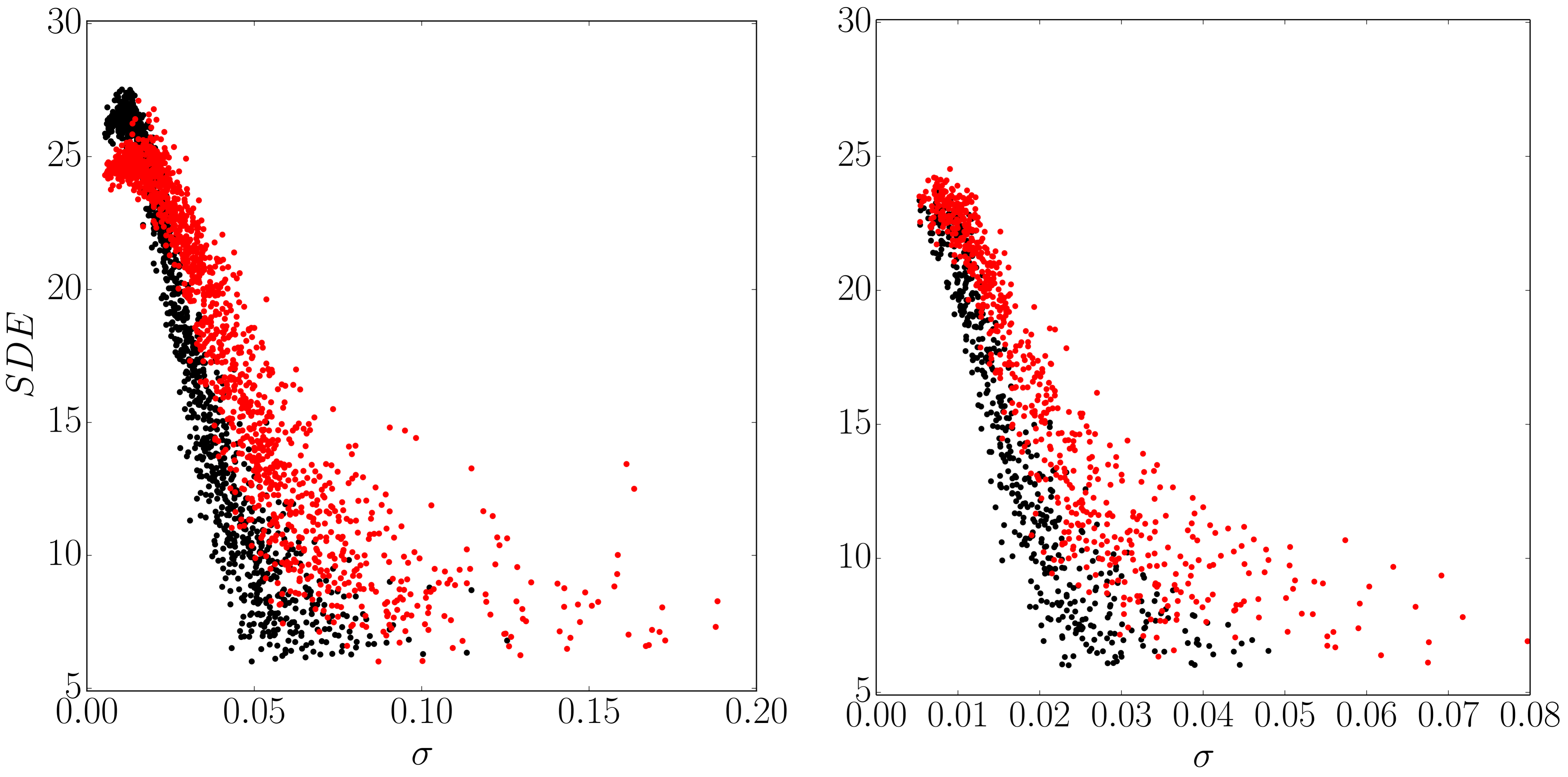}
    \caption{TFA vs TFAW detections. \textbf{Left:} SDEs of detections for Planet 1 in Table \ref{TransitParams} versus signal noise for TFA (black dots) and TFAW (red dots). \textbf{Right:} Same but for Planet 2 in Table \ref{TransitParams}.}
    \label{fig:TFAWdetect}
\end{figure}

\begin{table}[h]
\caption{Detection distributions for simulated planetary transits 1 and 2 as shown in Table~\ref{TransitParams} for three bins of total noise contributions.}
\label{tfawranges}
\centering
\resizebox{\linewidth}{!}{%
\begin{tabular}{ccllcllcll}
\hline\hline
\noalign{\smallskip}
 $\sigma$ & \multicolumn{3}{c}{$<0.01$} & \multicolumn{3}{c}{$0.01-0.04$} & \multicolumn{3}{c}{$>0.04$} \\ \hline
\multicolumn{1}{c|}{} & \multicolumn{1}{l}{$\mathrm{N}_\mathrm{TFA}$} & $\mathrm{N}_\mathrm{TFAW}$ & \multicolumn{1}{l|}{$\mathrm{N}_\mathrm{mut}$} & \multicolumn{1}{l}{$\mathrm{N}_\mathrm{TFA}$} & $\mathrm{N}_\mathrm{TFAW}$ & \multicolumn{1}{l|}{$\mathrm{N}_\mathrm{mut}$} & \multicolumn{1}{l}{$\mathrm{N}_\mathrm{TFA}$} & $\mathrm{N}_\mathrm{TFAW}$ & $\mathrm{N}_\mathrm{mut}$ \\ \cline{2-10} 
\multicolumn{1}{c|}{Planet 1} & - & \multicolumn{1}{c}{-} & \multicolumn{1}{c|}{75} & 76 & \multicolumn{1}{c}{131} & \multicolumn{1}{c|}{318} & 189 & \multicolumn{1}{c}{653} & \multicolumn{1}{c}{167} \\ \hline
\multicolumn{1}{c|}{} & \multicolumn{1}{l}{$\mathrm{N}_\mathrm{TFA}$} & $\mathrm{N}_\mathrm{TFAW}$ & \multicolumn{1}{l|}{$\mathrm{N}_\mathrm{mut}$} & \multicolumn{1}{l}{$\mathrm{N}_\mathrm{TFA}$} & $\mathrm{N}_\mathrm{TFAW}$ & \multicolumn{1}{l|}{$\mathrm{N}_\mathrm{mut}$} & \multicolumn{1}{l}{$\mathrm{N}_\mathrm{TFA}$} & $\mathrm{N}_\mathrm{TFAW}$ & $\mathrm{N}_\mathrm{mut}$ \\ \cline{2-10} 
\multicolumn{1}{c|}{Planet 2} & 10 & \multicolumn{1}{c}{41} & \multicolumn{1}{c|}{107} & 67 & \multicolumn{1}{c}{393} & \multicolumn{1}{c|}{491} & 2 & \multicolumn{1}{c}{69} & \multicolumn{1}{c}{14} \\ \hline
\end{tabular}%
}
\end{table}

\begin{figure}[bth]
    \centering
    \includegraphics[width=\linewidth,height=10.0cm, keepaspectratio]{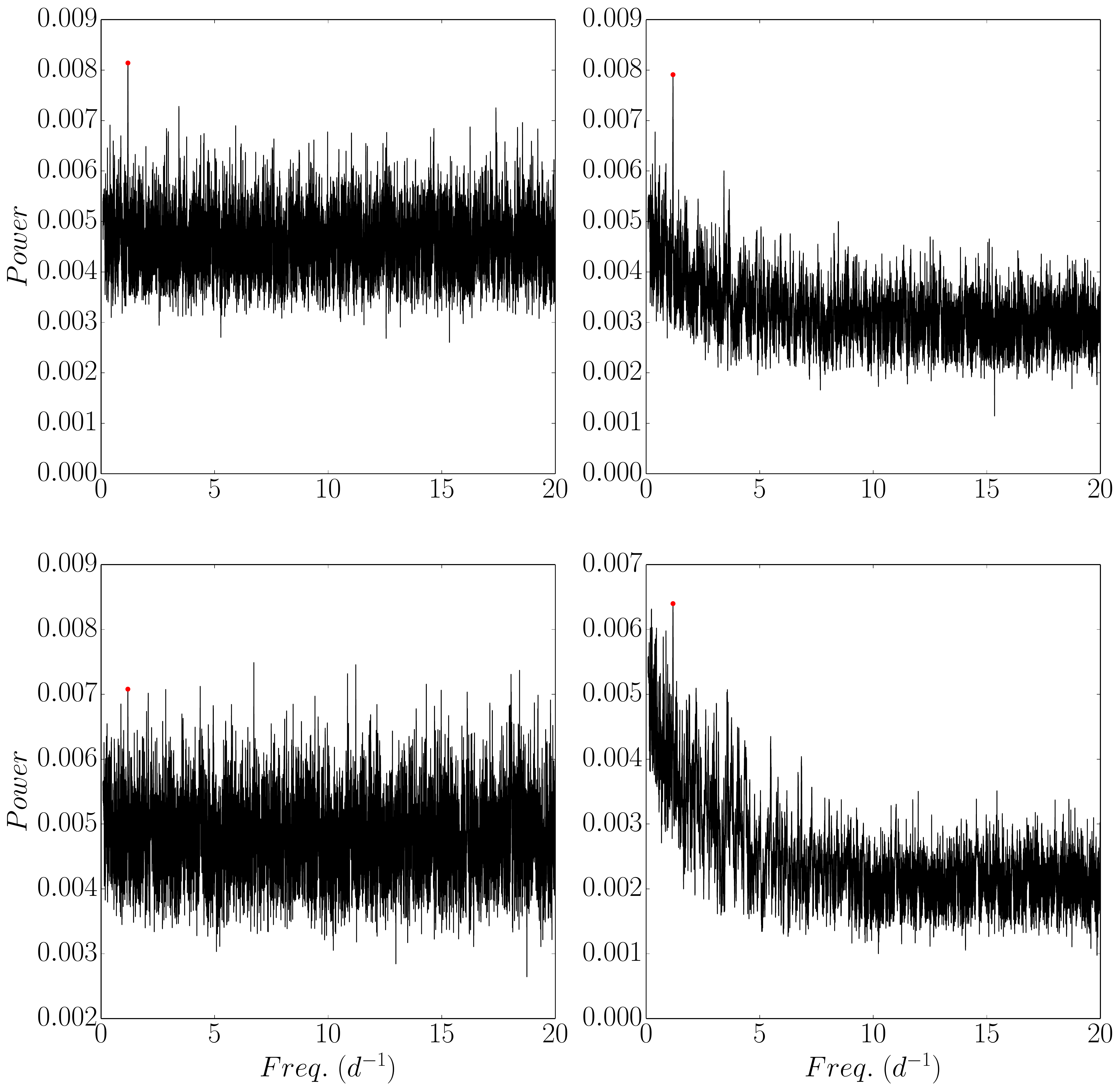}
    \caption{TFA vs TFAW BLS power spectrum examples for Planet 1 in Table \ref{TransitParams}. \textbf{Top:} TFA BLS power spectrum (left) vs TFAW BLS power spectrum (right) for a simultaneous detection as defined in Sect. \ref{sect:transitdetect}. Red dots mark the true period of the transit. \textbf{Bottom:} Same but for a TFAW mutually exclusive detection.}
    \label{fig:TFAWpwscomp}
\end{figure}

The example shown in Table~\ref{tfawdetect} shows that the number of non-simultaneous detections (both in absolute and percentage) is always higher (a factor $\sim$3$\times$ for Planet 1 and $\sim$6$\times$ for Planet 2) in the case of TFAW light curves for both planetary transits. Also the mean SDE values are higher for TFAW than for TFA. As can be seen in Fig.~\ref{fig:TFAWdetect} the SDE improvement for a given light curve can be up to a factor $\sim$2.5$\times$ for low SNR signals. Table \ref{tfawranges} shows the distribution of the TFA and TFAW detections in three total noise contribution bins. 
In the high-SNR regime both TFA and TFAW behave in a similar way. TFAW performs better than TFA in detecting the transit signal in the mid-SNR regime ($\sigma$=0.01-0.04) by a factor $\sim$1.16$\times$ for Planet 1 and a factor $\sim$1.6$\times$ for Planet 2. In the low-SNR case ($\sigma>$0.04), TFAW detects the transit in $\sim$2.3$\times$ more light curves than TFA for Planet 1 and in $\sim$5.2$\times$ for Planet 2. Taking into account the whole noise range, TFAW improves the detection rate by a factor $\sim$1.6$\times$ for both planetary transits. Figure \ref{fig:TFAWpwscomp} shows two examples of the increase in the SDE of the BLS power spectrum peaks for Planet 1 in Table \ref{TransitParams} in the low SNR regime. Top two panels correspond to a simultaneous detection for which TFAW yields an increased SDE for the true period of 11.2 versus 6.2 for TFA. The bottom two panels present the power spectrum for a TFAW mutually exclusive detection. In this case, the peak corresponding to the true period of the transit is hidden within the noise peaks of the TFA power spectrum with a SDE of $\sim$3.9 (below the detection threshold defined above). For TFAW, the peak can easily be identified and has a SDE of $\sim$8.3 (a factor $\sim$2.1$\times$ higher than for TFA). In addition, TFAW is able to detect the planetary signals, with a SDE above 6, up to a total noise of $\sim$0.19 mag for planet 1 and up to $\sim$0.07 mag for planet 2 whereas TFA is able to detect them up to $\sim$0.1 and $\sim$0.045 mag respectively.

\subsection{TFAW vs TFA signal reconstruction}
\label{sect:signalrec}

To illustrate the noise and trend filtering efficiency of the TFAW method during the iterative signal reconstruction step, we generate a sinusoidal signal affected by a combination of different levels of simulated Gaussian, $\sigma_\mathrm{w}$, and correlated, $\sigma_\mathrm{r}$ noises. As in Sect. 3.1 we also introduce linear and exponential trends, as well as gaps and jumps in the data. The period of the simulated signal is 0.63 days and 0.03 mag amplitude. Again, we use a template of 250 simulated template stars suffering from the same trends and systematics as the target star, each of them affected by a random distribution of $\sigma_\mathrm{w}$ and $\sigma_\mathrm{r}$ combinations (see Table~\ref{NoiseParams} for details).

   \begin{table}[h]
   \caption{Noise parameters for the simulated light curves}           
   \label{NoiseParams}      
   \centering                         
   \begin{tabular}{c c c c}       
   \hline\hline
   \noalign{\smallskip}
   Signal type & SNR & $\sigma_\mathrm{w}$ (mag) & $\sigma_\mathrm{r}$ (mag)\\     
   \hline                       
   \noalign{\smallskip}
   Sinusoidal & High & 0.01 & 0.005 \\     
   Sinusoidal & Low & 0.1 & 0.005 \\     
   Transit & High & 0.01 & 0.005 \\     
   Transit & Low & 0.04 & 0.005 \\     
   \hline                                   
   \end{tabular}
   \end{table}

\begin{figure}[h]
    \centering
    \includegraphics[width=\linewidth,height=10.0cm, keepaspectratio]{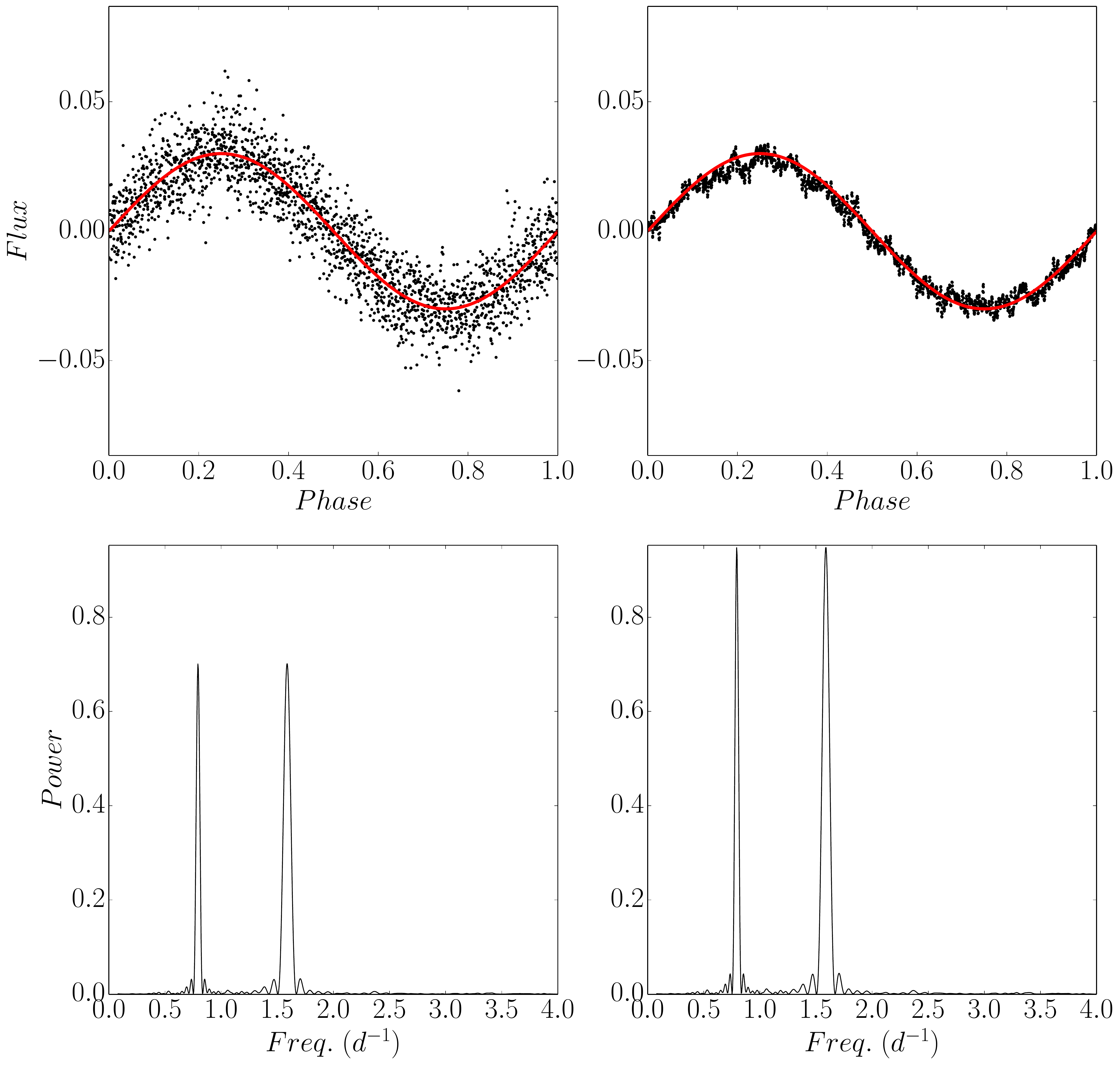}
   \caption{Noise filtering comparison of a simulated sinusoidal signal. The same number of template stars and LS parameters were used for TFA and TFAW. \textbf{Top left:} TFA-detrended and reconstructed phase folded signal. Red line corresponds to the simulated signal. \textbf{Top right:} The same phase folded signal but TFAW-detrended, reconstructed and de-noised. \textbf{Bottom left:} LS power spectrum of TFA frequency analysis step. \textbf{Bottom right:} LS power spectrum of TFAW frequency analysis step.}

   \label{PureSinusoidalln}
   \end{figure}

Fig.~\ref{PureSinusoidalln} shows the noise filtering capabilities of TFAW compared to TFA for a simulated target star with a sinusoidal modulation. While the noise dispersion is clearly diminished, there is no modification to either the amplitude or the phase of the signal, and the sampling timescale of the signal remains unchanged (unlike in simple binning). We also show the LS power spectrum obtained in the frequency analysis step. As can be seen the height and amplitude of the peaks for the TFAW case have a higher power value compared to the TFA ones.

   \begin{figure}[h]
    \centering
    \includegraphics[width=\linewidth,height=10.0cm, keepaspectratio]{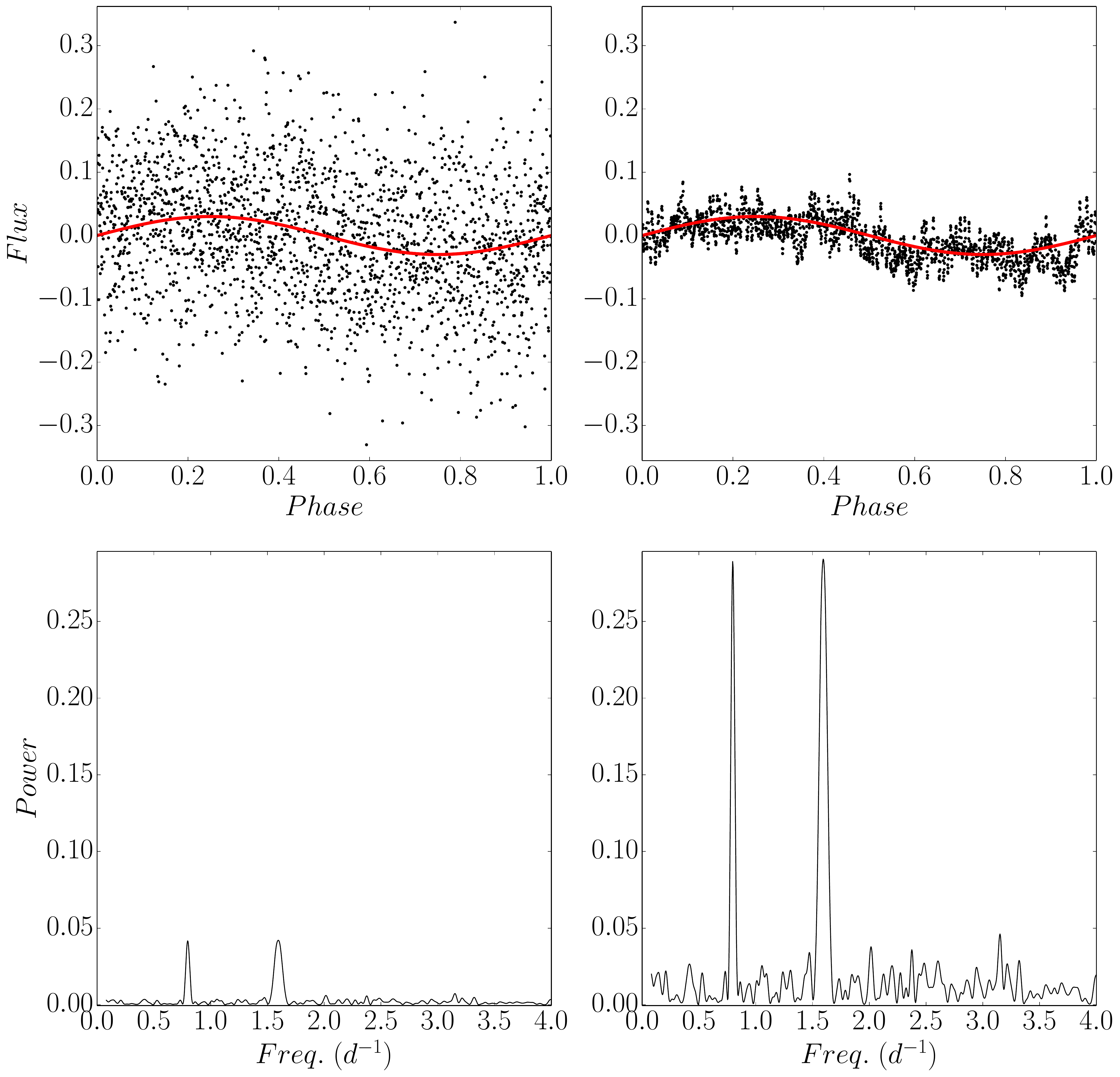}
   \caption{Noise filtering comparison of a simulated sinusoidal signal with lower SNR. Same notation, TFA and TFAW parameters as Fig.~\ref{PureSinusoidalln}.}

   \label{PureSinusoidalhn}
   \end{figure}

In Fig.~\ref{PureSinusoidalhn}, the same sinusoidal signal is simulated but with lower SNR. As before, the noise dispersion is also significantly diminished while the period and amplitude of the signal remains unchanged. In this case, the LS power spectrum is improved by a greater factor than the one for the high SNR case. However, in this case, although the amplitude of the power spectrum peaks is greater in the case of TFAW, the SDE values for both TFA and TFAW are very similar (10.55 and 10.6, respectively) due to the low LS continuum.

   \begin{figure}[h]
    \centering
    \includegraphics[width=\linewidth,height=10.0cm, keepaspectratio]{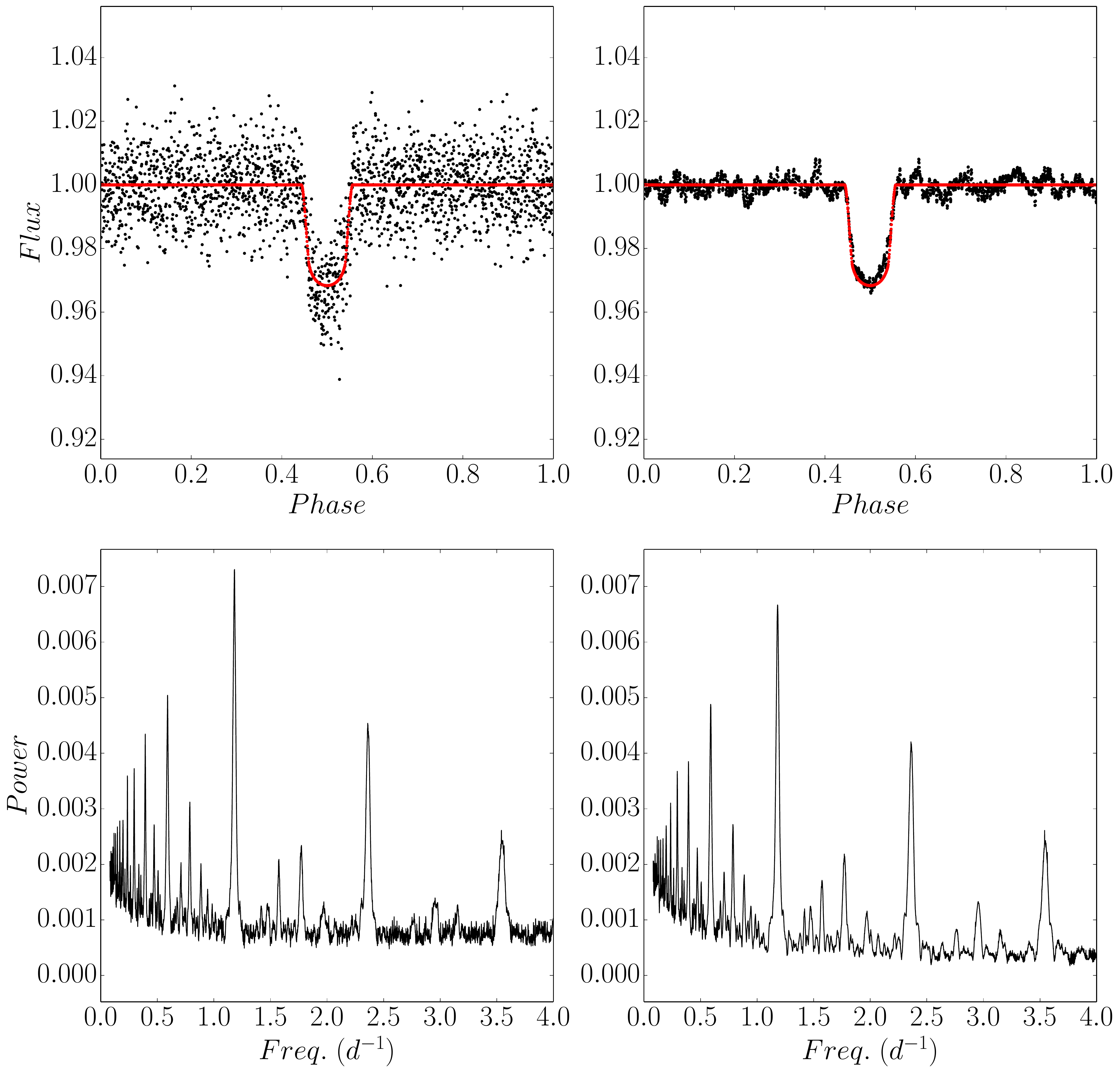}
   \caption{Noise filtering comparison of a simulated box-shaped transit (Planet 1 in Table \ref{TransitParams}). Same notation, TFA and TFAW parameters as Fig.~\ref{PureSinusoidalln}.}
   \label{Transitln}
   \end{figure}

   \begin{figure}[h]
    \centering
    \includegraphics[width=\linewidth,height=10.0cm, keepaspectratio]{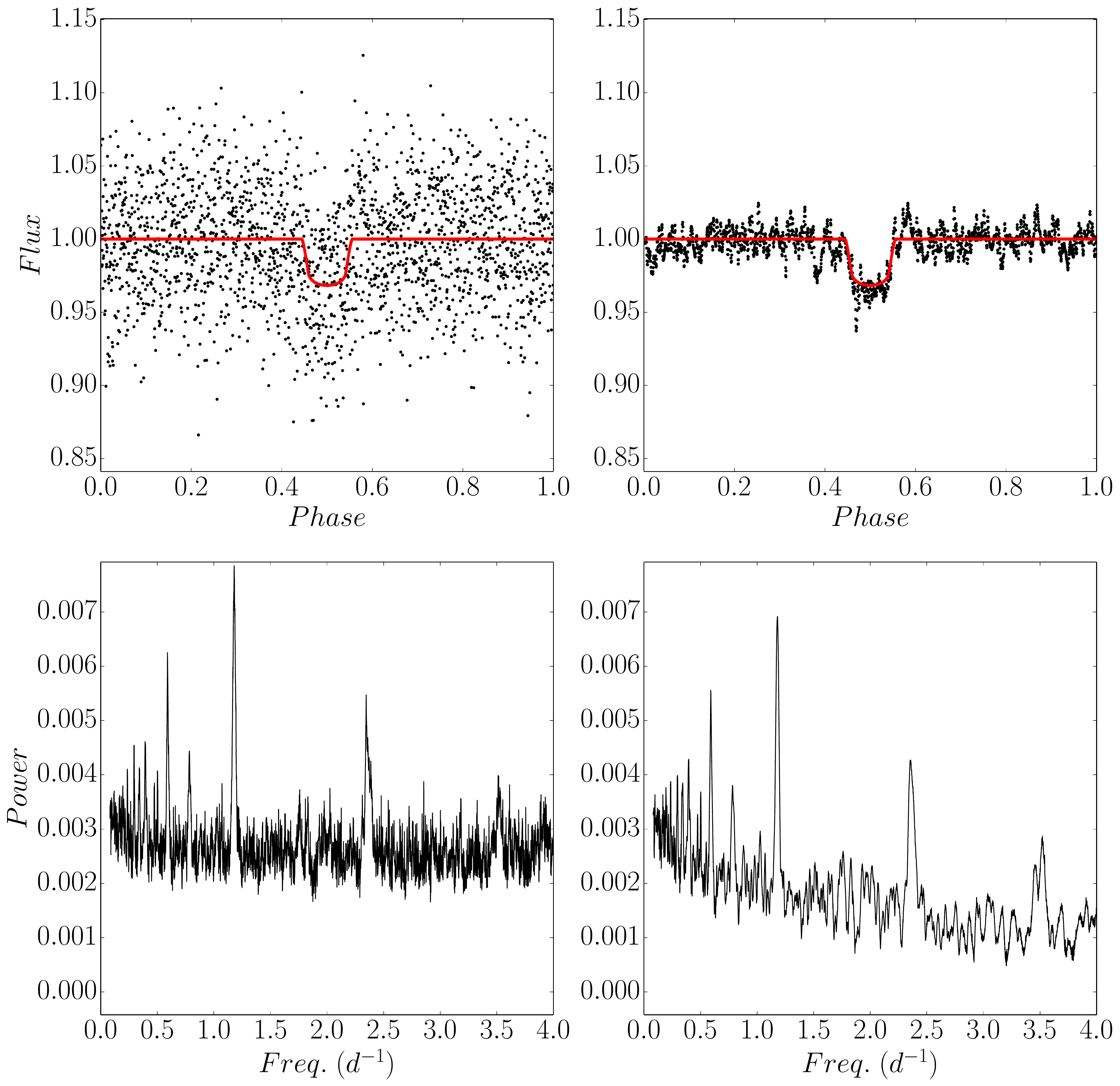}
   \caption{Noise filtering comparison of the same simulated box-shaped transit as in Fig.~\ref{Transitln} with lower SNR. Same notation, TFA and TFAW parameters as Fig.~\ref{PureSinusoidalln}.}
   \label{fig:Transithn}
   \end{figure}

Fig.~\ref{Transitln} simulates a planetary transit. We use {\tt batman}, with the parameters of Planet 1 in Table~\ref{TransitParams} to simulate a~\citet{Mandel2002} planetary transit model, and the noise parameters in Table~\ref{NoiseParams}. The high-frequency noise is filtered by TFAW during the iterative reconstruction step, resulting in a better defined transit without modifying its shape and depth. This result overcomes the artificial ripples and transit depth modification introduced in Fig. 6 of~\citet{Grziwa2016a} due to wavelet coefficient thresholding. 

The TFAW BLS power spectrum for this high SNR planetary transit remains almost unchanged with respect to the TFA version (with a lower noise floor), because the BLS power spectrum is not primarily limited by high-frequency random noise. The SDE values for this case are also very similar; 23.6 for TFA and 24.5 for TFAW.

Fig.~\ref{fig:Transithn} shows the same transit but with a much larger noise component (see Table \ref{NoiseParams}). As in the case of the simulated sinusoidal with low SNR (see Fig.~\ref{PureSinusoidalhn}), TFAW improves the characterization of the signal's shape while diminishing the noise, recovers the correct period and transit depth, and improves the detectability of the transit signal in the BLS periodogram (with and SDE of 17.8 for TFAW and 13.3 for TFA).

\subsection{Application to multiperiodic signals}
\label{sect:multi}

\begin{figure}[h]
    \centering
    \includegraphics[width=\linewidth,height=20.0cm, keepaspectratio]{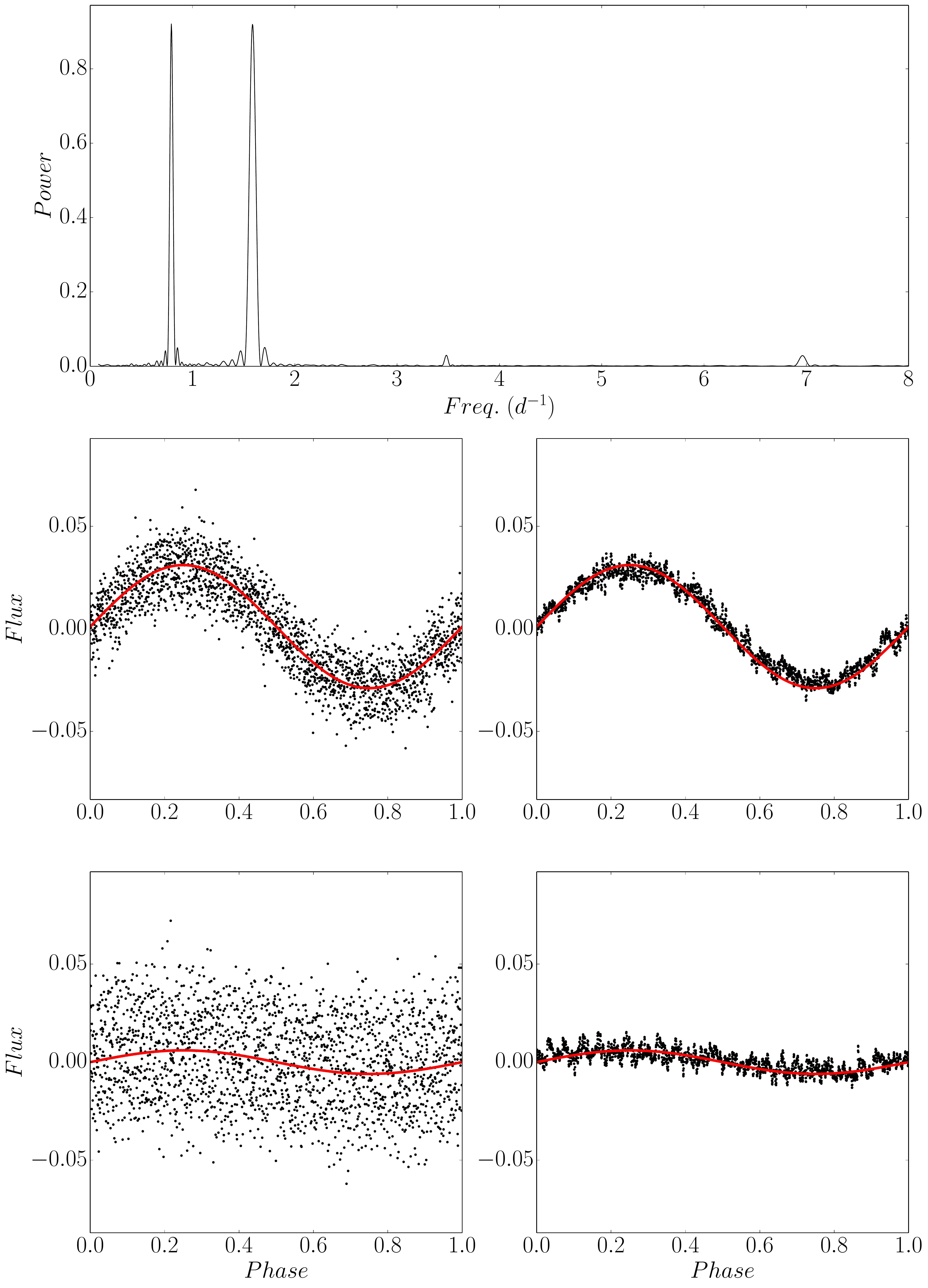}
   \caption{Example of the signal recovery for a multi-periodic sinusoidal signal. Red line corresponds to simulated signal. \textbf{Top:} LS power spectra of the signal after TFAW frequency analysis step (notice the small peak around 6.9 d$^{-1}$ corresponding to the secondary
signal). \textbf{Middle left:} TFA-detrended and reconstructed phase-folded low frequency signal. \textbf{Middle right:} Same TFAW-filtered phased folded low frequency signal. \textbf{Bottom left:} TFA-detrended and reconstructed secondary signal. \textbf{Bottom right:} TFAW phase-folded secondary signal.}
   \label{TwoSinusoidals}
   \end{figure}

~\citet{Kovacs2008} presents a variation of TFA to extend its application to multiperiodic signals using their Fourier representation (for the cases in which such representation is adequate). However, they state that if the signal has additional components such as transients or planetary transits, a more complicated model should be used to approximate each of those extra signals present in the light curve. TFAW can be used to separate the different signal contributions directly from the signal after TFA's frequency analysis step (provided that the periods of each of the signals have been previously and correctly determined using BLS or another method). As a first example, we simulate the same high SNR-type sinusoidal signal as in Fig.~\ref{PureSinusoidalln} and Fig.~\ref{PureSinusoidalhn} modulated with a 0.1438-day sinusoid with an amplitude of 0.006 mag, and with noise parameters according the first row of Table~\ref{NoiseParams}. As can be seen in Fig.~\ref{TwoSinusoidals}, if we are able to find the correct frequency of the secondary signal (notice the peak around 6.9 d$^{-1}$ in the corresponding LS periodogram), we can fully recover it by applying TFAW directly (i.e. with no need of subtracting the primary signal)  to the raw data without modifying the amplitude and shape of any of the underlying signals.

\begin{figure}[h]
    \centering
    \includegraphics[width=\linewidth,height=20.0cm, keepaspectratio]{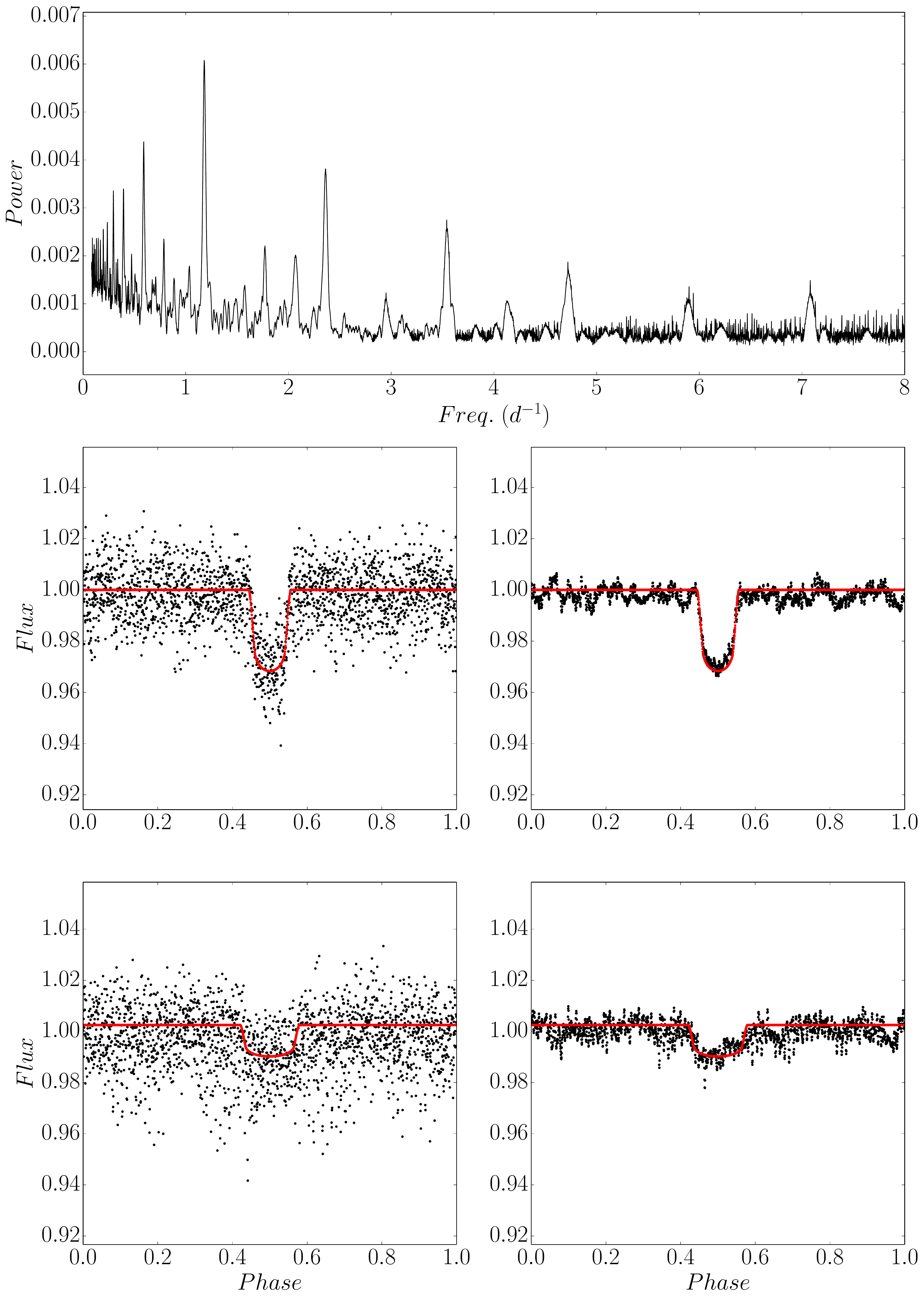}
   \caption{Example of the signal recovery in a multi-transit light curve.  Red line corresponds to simulated signal. \textbf{Top:} BLS power spectra of the signal after TFAW frequency analysis step. Check the peak around 2 d$^{-1}$ corresponding to the secondary transit. \textbf{Middle left:} TFA-detrended and reconstructed phase-folded Planet 1 signal. \textbf{Middle right:} Same TFAW-filtered phased folded Planet 1 signal. \textbf{Bottom left:} TFA-detrended and reconstructed Planet 2 signal. \textbf{Bottom right:} TFAW phase-folded Planet 2 signal.}
   \label{TwoTransits}
   \end{figure}

The second example in Fig.~\ref{TwoTransits} shows the results of applying TFAW to a high SNR-type light curve affected by two planetary transits simulated with {\tt batman} using the parameters of planets 1 and 2 in Table~\ref{TransitParams}, and noise parameters according the second-to-last row of Table~\ref{NoiseParams}. Again, if we are able to find the period of the secondary transit after the frequency analysis step, we can separate the secondary transit from the primary simply by phase folding the raw light curve and applying TFAW. As in the other examples, we are also able to improve the SNR of both the primary and secondary signals.   

\subsection{Wavelet versus bin average signal approximation}
\label{sect:binvswave}

\begin{figure}[h]
    \centering
    \includegraphics[width=\linewidth, keepaspectratio]{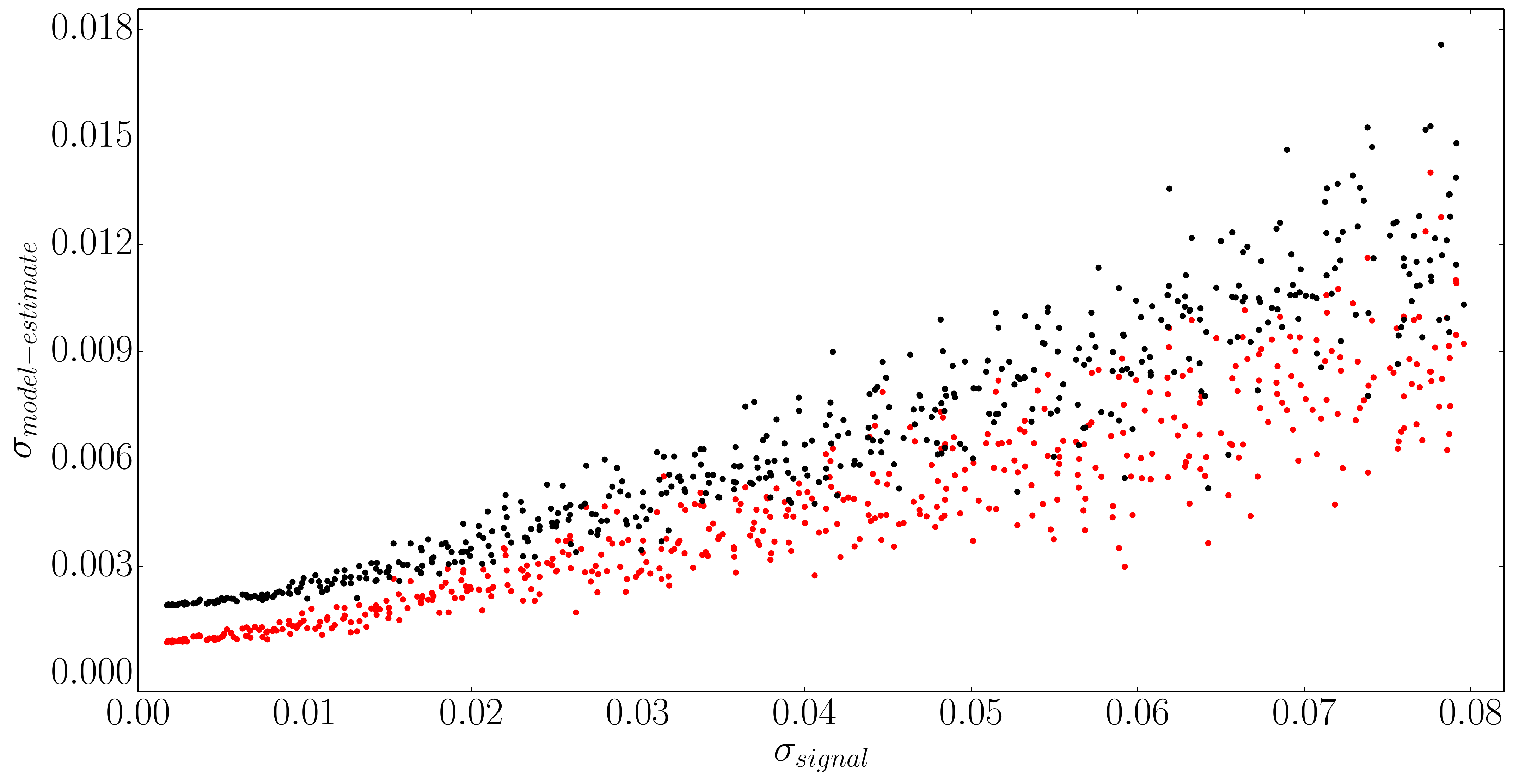}
    \caption{Wavelet signal approximation versus bin average comparison. Comparison of the standard deviations of the estimated signals obtained by the sum of the ISWTs given by \textit{signal level} (red) and the one given by bin averaging (black) for Planet 1 in Table \ref{TransitParams} for decreasing transit depth.}
	\label{fig:model_vs_estimate}
	\end{figure}

In order to test the likeness of the signal approximation, $\{A(i)\}$, to the inserted signal TFAW compared to the bin averaging used by TFA, we use Planet 1 in Table \ref{TransitParams}. We inject it in 500 light curves, using {\tt batman} and a~\citet{Mandel2002} planetary transit model, for increasing values of noise contribution (i.e. lower transit depth). For each simulation, we compute the signal estimation given by the SWT and bin averaging with 100 bins and obtain their deviation with respect to the simulated input signal. Fig.~\ref{fig:model_vs_estimate} shows that, in general, the estimation of the transit shape given by the sum of the ISWTs defined by \textit{signal level} provides a better representation than the one given by the TFA bin average method. This is especially true for low $\sigma_\mathrm{signal}$, with up to a factor of $\sim$2$\times$ of improvement.

\begin{figure}[h]
   \centering
   \includegraphics[width=\linewidth,height=10.0cm, keepaspectratio]{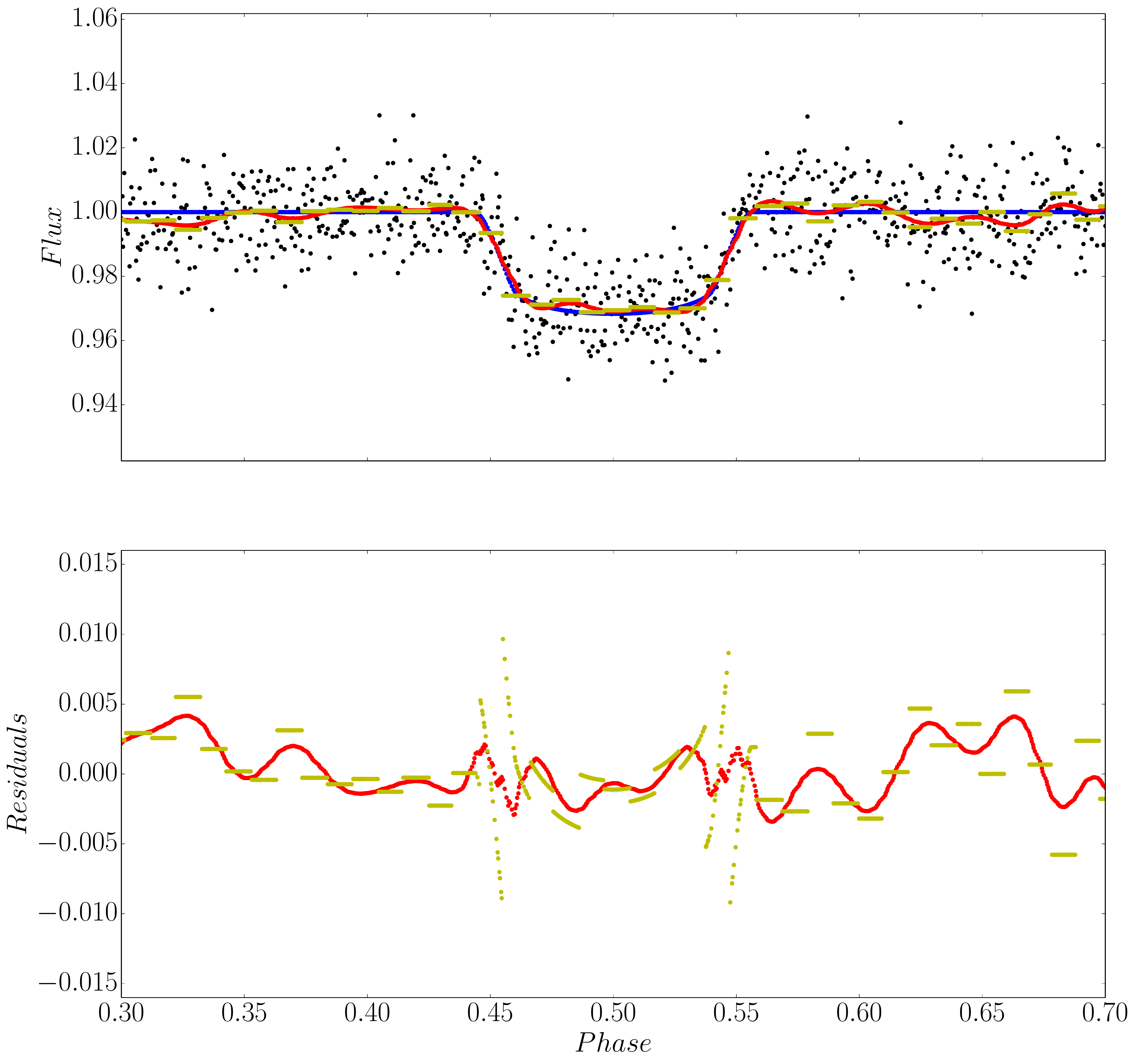}
   \caption{Wavelet signal approximation versus bin average comparison for Planet 1 in Table \ref{TransitParams}. \textbf{Top}: Planetary transit with high SNR. Blue line corresponds to the input~\citet{Mandel2002} model, red line represents the wavelet approximation of the signal, and yellow line is the bin average approximation. \textbf{Bottom}: Difference between the planetary model, and the wavelet and bin average approximations (same color notation).}
   \label{fig:model_vs_estimate_transit}
   \end{figure}

In the case of planetary transits, as can be seen in Fig.~\ref{fig:model_vs_estimate_transit}, the improvement in the signal approximation is due to the fact that the wavelet reconstruction of the signal better fits the ingress and egress profiles of the transit even in the cases of low signal-to-noise ratios compared to the bin average method.

\subsection{Comparison of TFA and TFAW transit parameters fit values and uncertainties}
\label{sect:MCMC}
We want to compare TFA vs. TFAW performance in terms of assessing the bias of the fitted transit parameters values and their uncertainties. To quantify those, a low SNR-type planetary transit of the Planet 1 according Table~\ref{TransitParams}~and last row of Table~\ref{NoiseParams} was considered. This was modeled with {\tt batman}, following a~\citet{Mandel2002} analytic transit model. We used the Markov Chain Monte Carlo (MCMC) sampler provided by {\tt emcee}\footnote{\url{http://dfm.io/emcee}}~\citep{Foreman2013} to sample the posterior distribution of the 6 transit parameters (a, q, i, per, p, l). Keeping the eccentricity fixed, we consider a uniform distribution for the priors and run the sampler with 200 chains and 5000 iterations with a burn-in phase of 1000 iterations.

\begin{figure*}
    \centering
    \begin{subfigure}[t]{0.5\linewidth}
        \centering
        \includegraphics[height=9cm]{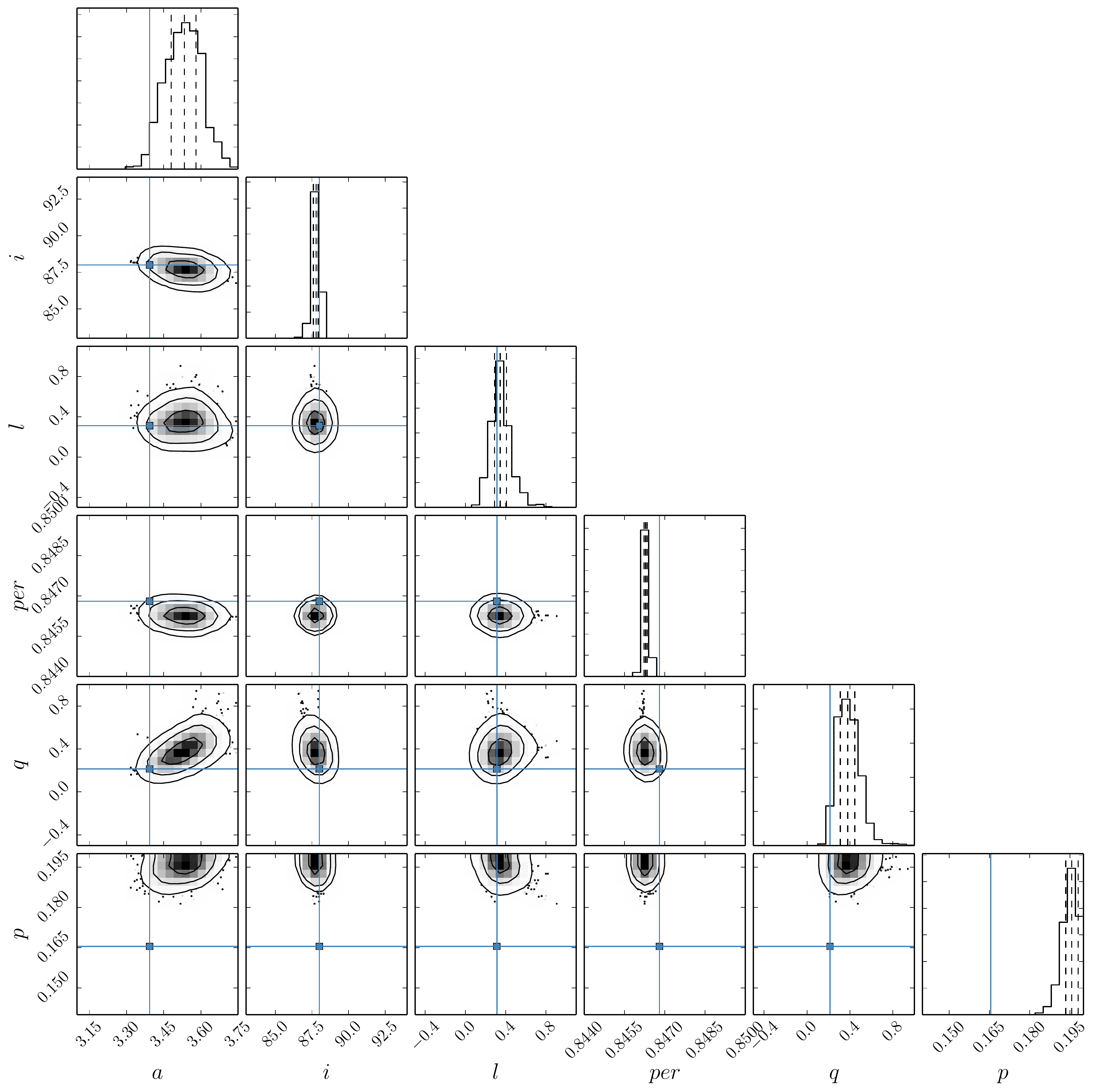}
    \end{subfigure}%
    ~ 
    \begin{subfigure}[t]{0.5\textwidth}
        \centering
        \includegraphics[height=9cm]{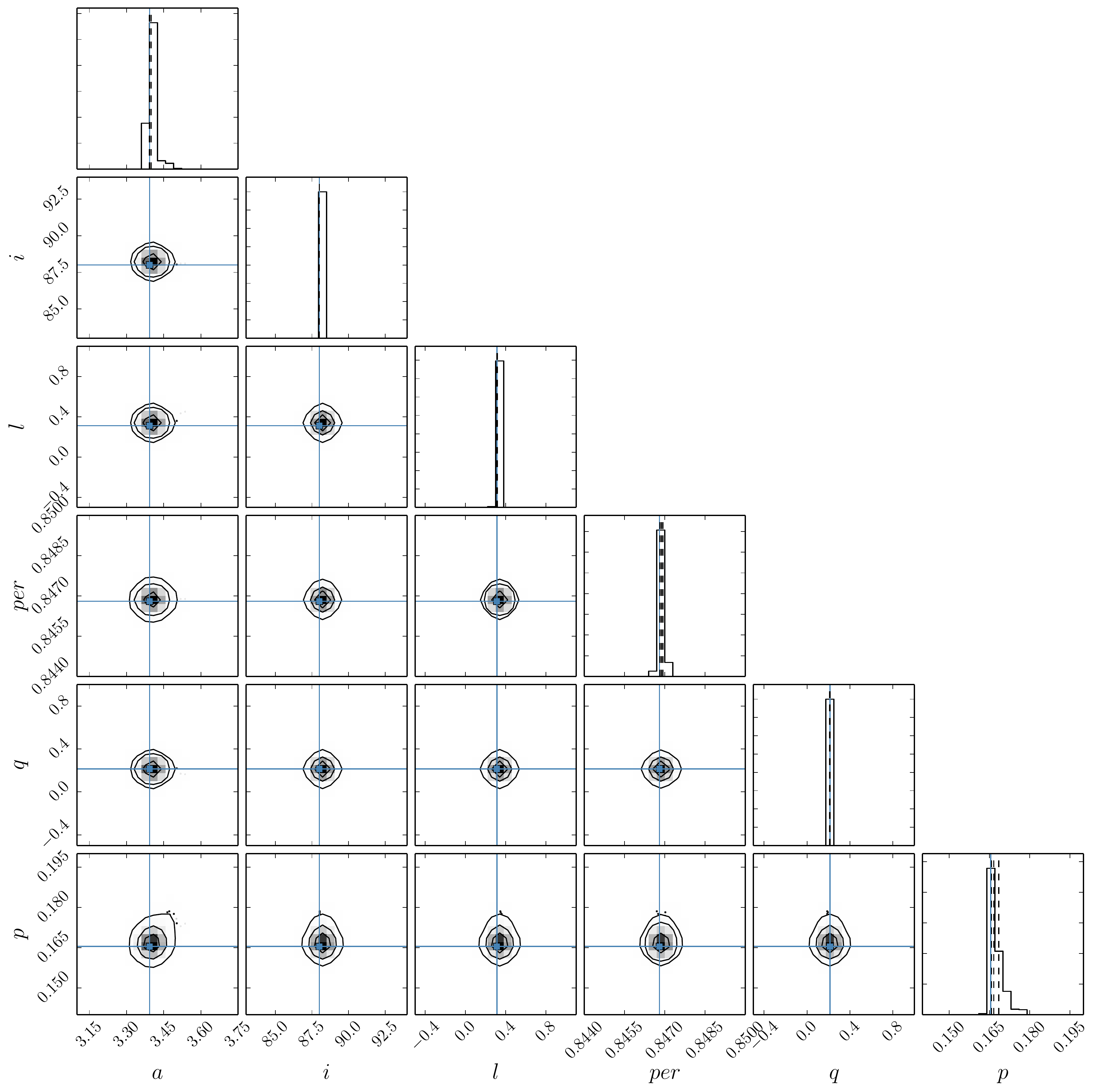}
    \end{subfigure}
      \caption{1-D and 2-D projections of the posterior probability distributions of the 6 MCMC fitted parameters for the TFA (left) and TFAW (right) detrended light curves. The injected values for (a, q, i, P, p, l) are marked in solid blue. The  25\%, 50\%, 75\% quantiles, are displayed in dash vertical lines on the 1-D histograms.}
   \label{fig:banana_plots}
   \end{figure*}

Fig.~\ref{fig:banana_plots} shows the 1-D and 2-D projections of the posterior probability distributions of the 6 MCMC fitted parameters for the TFA and TFAW detrended light curves. Similarly, at top panel of Table~\ref{TransitParamsMCMC} we compare the injected transit parameters values with the ones obtained through MCMC for TFA and TFAW posterior probability distributions. MCMC parameter values correspond to the 50\% quantile while the uncertainties are computed from the 25\% and 75\% quantiles as the upper and lower errors. In the case of TFA, while some of the distributions are fairly behaved, for parameters a, q and p, they present wider features either characterized by larger uncertainties (a, q) and/or larger biases (a, p). On the other hand, for TFAW light curves, the biases with respect to the initial parameters values are strongly diminished compared to TFA ones. As for the uncertainties, they are, in general, smaller for TFAW than for TFA. In some parameters, such as i, l and a, they are largely decreased. At the bottom panel of Table~\ref{TransitParamsMCMC}, the 95\% confidence highest probability density credibility intervals for both TFA and TFAW MCMC cases, are shown. The width of the credibility intervals is $\sim$10$\times$ narrower (except for the period which is better determined) in the case of TFAW compared to the TFA ones.

   \begin{table*}
   \caption{Top table: Actual parameters values used to simulate the transit of Planet 1. Posterior transit parameters values and their uncertainties (with the 25\% and 75\% quantile as the upper and lower errors) for TFA and TFAW MCMC fits. Bottom table: 95\% confidence intervals of the highest probability density for Planet 1 transit parameters TFA and TFAW MCMC fits.}   
   \label{TransitParamsMCMC}      
   \centering                         
   \begin{tabular}{c c c c c c c}       
   \hline\hline
   \noalign{\smallskip}
   Parameters & a ($R_{*}$) & q & i ($^{\circ}$) & per (days) & p & l\\   
   \hline
   \noalign{\smallskip}
   Simul. values & 3.392862 & 0.214 & 88 & 0.8468 & 0.16542286 & 0.312 \\ 
   \noalign{\smallskip}   
   TFA MCMC & 3.53363$^{+0.06460}_{-0.08075}$ & 0.37952$^{+0.10736}_{-0.09784}$ & 87.7895$^{+0.19976}_{-0.28489}$ & 0.84628$^{+0.0009}_{-0.00008}$ & 0.19564$^{+0.00347}_{-0.0032}$ & 0.34398$^{+0.09770}_{-0.08413}$ \\
   \noalign{\smallskip}
   TFAW MCMC & 3.39438$^{+0.01072}_{-0.00228}$ & 0.21317$^{+0.00206}_{-0.00591}$ & 88.0075$^{+0.00687}_{-0.00212}$ & 0.84687$^{+0.00008}_{-0.0001}$ & 0.16656$^{+0.00312}_{-0.00102}$ & 0.31259$^{+0.00603}_{-0.00189}$ \\
   \noalign{\smallskip}
   \hline                                   
   \noalign{\smallskip}
   & \multicolumn{6}{c}{95\% confidence intervals of the highest posterior density}\\ 
   \noalign{\smallskip}
   \hline    
   \noalign{\smallskip}
   TFA MCMC & 3.368 - 3.672 & 0.186 - 0.577 & 87.31 - 88.21 & 0.8461 - 0.8465 &  0.186 - 0.204  & 0.161 - 0.564 \\
   \noalign{\smallskip}
   TFAW MCMC & 3.384 - 3.432 & 0.198 - 0.219 & 87.99 -  88.02 & 0.8467 - 0.8470 & 0.164 - 0.172 & 0.305 -  0.335\\
   \noalign{\smallskip}
   \hline                                   
   \end{tabular}
   \end{table*}

\section{Application to Real Survey Light Curves}
In this section we test the TFAW algorithm on real data from the wide FoV ground-based TFRM-PSES survey.

\subsection{The Telescope Fabra-ROA at Montsec}
\label{sect:tfrm}
The Telescope Fabra-ROA at Montsec (TFRM) is a refurbished f/0.96 0.50m Baker-Nunn Camera~\citep{Henize1957} for robotic CCD wide-FoV surveying purposes~\citep{Fors2013}. The project is a joint collaboration between the Reial Acad{\`e}mia de Ci{\`e}ncies i Arts de Barcelona - Observatori Fabra and the Real Instituto y Observatorio de la Armada. The telescope is installed at the summit of the Montsec d'Ares (Spain) as part of the Observatori Astron{\`o}mic del Montsec. 

The TFRM is conducting a targeted M-dwarfs transiting exoplanetary survey (TFRM-PSES) since its deployment at the end of 2011.
TFRM-PSES benefits from the wide FoV of 19.4 sq. deg. for a 16\,MPix camera, to deliver at 25\,sec-cadence few-mmag precision-level light curves for every object brighter than R=14\,mag~\citep{delSer2015,Fors2013}.
 
\subsubsection{Data description}
\label{subsect:tfrmdata}
The data from the TFRM-PSES survey used for the TFAW performance assessment comprises 2048 data points from 30 nights observed during 2013, 2014 and 2015 for a field centered at $(\alpha,~\delta)$ = (10:14:44, +48:30:00). All light curves in the data set have been observed with a 475\,nm cutoff frequency glass filter (Schott GG475), 17\,sec exposure time and an altitude above 50$^{\circ}$. The light curves were generated with the APEX-2 reduction pipeline~\citep{Devyatkin2010}. The pipeline includes aperture photometry, atmospheric extinction correction, differential photometry (using the BESTRED algorithm ~\citep{Voss2006}) and astrometric calibration.

\subsubsection{Results}
\label{subsect:tfrmresults}
\begin{figure}[h]
    \centering
    \includegraphics[width=\linewidth, keepaspectratio]{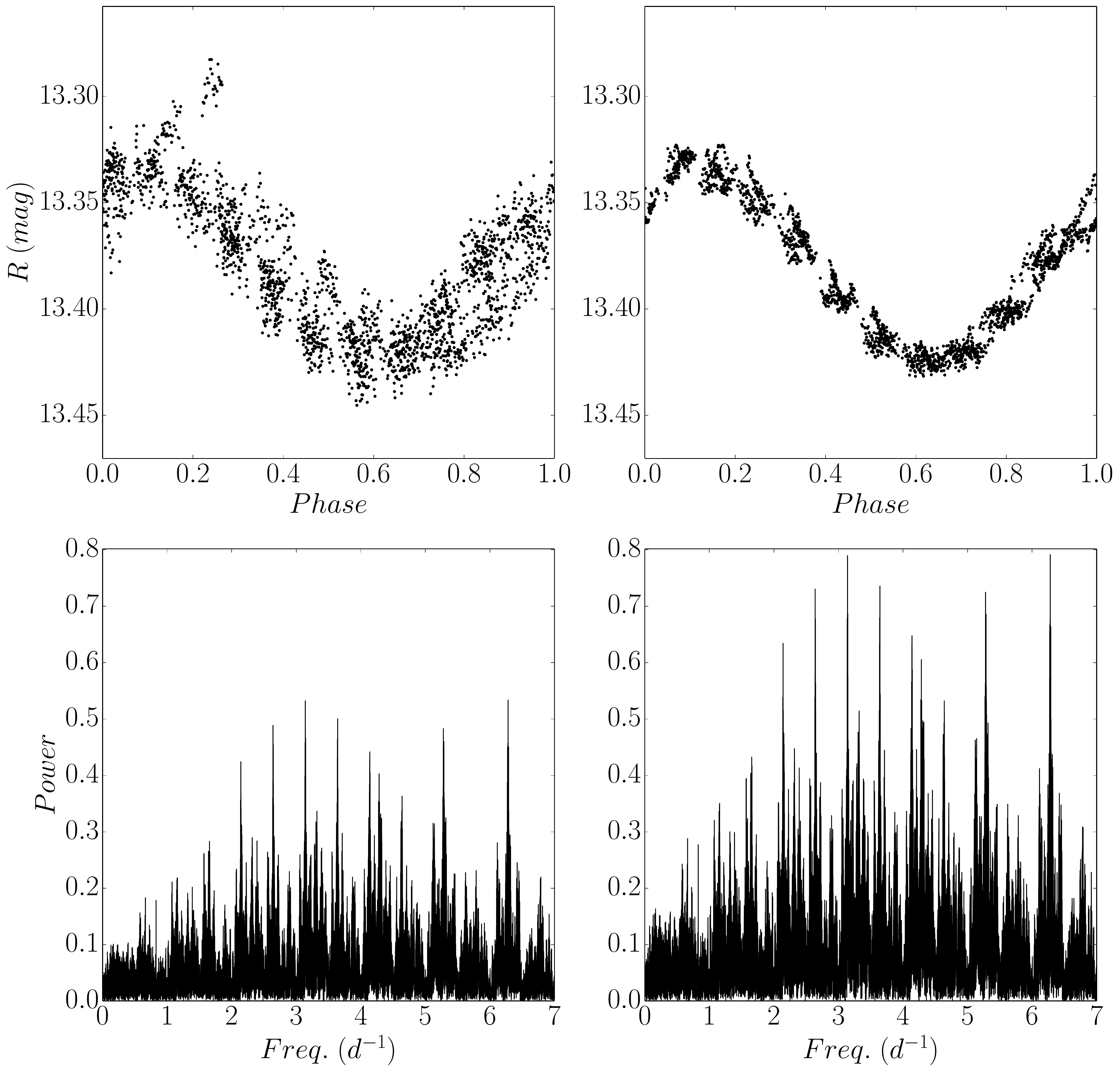}   
   \caption{Example of the TFAW filtering capabilities on the observed TFRM-PSES light curve of \object{2MASS J10144313+5018191}, a newly-discovered variable star. Same notation and TFAW parameters as Fig.~\ref{PureSinusoidalln}.}
   \label{fig:TFRM1}
   \end{figure}

\begin{figure}[h]
    \centering
    \includegraphics[width=\linewidth, keepaspectratio]{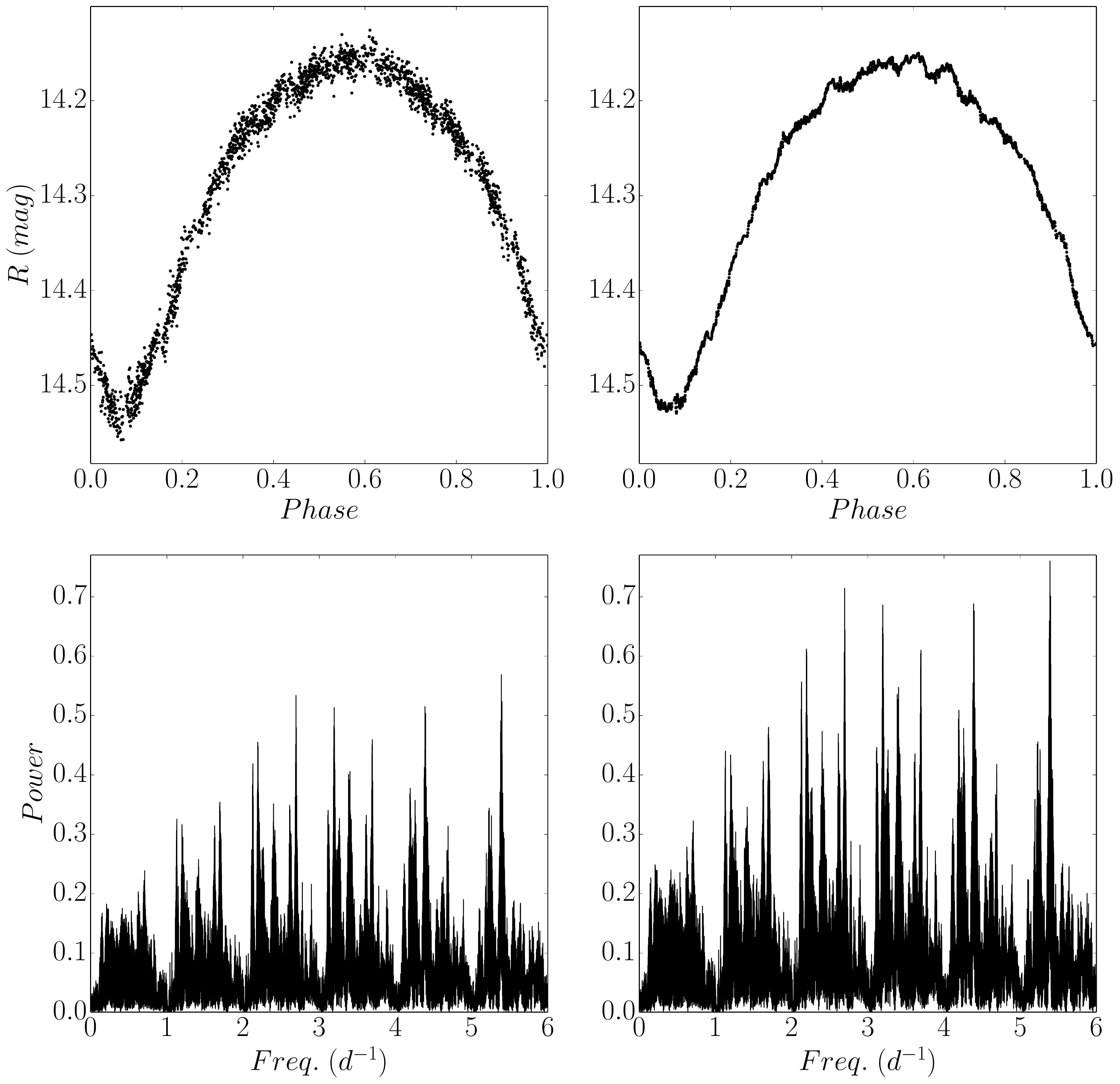}
   \caption{Example of the TFAW filtering capabilities on the observed TFRM-PSES light curve of \object{NSVS 4921994}. Same notation and TFAW parameters as Fig.~\ref{PureSinusoidalln}.}
   \label{fig:TFRM2}
   \end{figure}

In Fig.~\ref{fig:TFRM1} and Fig.~\ref{fig:TFRM2} we present two examples obtained from the TFRM-PSES data. Fig.~\ref{fig:TFRM1}, \object{2MASS J10144313+5018191}, is a newly-discovered variable star, detected both with TFA and TFAW, probably a $\delta$-Scuti star with a 0.1592-day period. As can be seen, after TFAW is applied, the noise contribution decreases significantly and outliers are efficiently removed. The LS power spectrum during the frequency analysis step greatly improves with respect the TFA one presenting higher power of the peaks and an increase in the SDE (10.1 for TFAW in front of 9.8 for TFA).
The one in Fig.~\ref{fig:TFRM2}, a W Ursae Majoris-type variable star with a 0.371018-day period (the maximum peak in the LS power spectrum corresponds half the cataloged period) is cataloged as \object{NSVS 4921994}~\citep{Wozniak2004}. As in the previous case, the SNR of the signal is improved and the correct period is recovered from the TFAW LS power spectrum. Also, the power spectrum itself presents higher peaks and an increased SDE (10.31 for TFAW in front of 10.1 for TFA). 

The small increase in the SDEs of both example light curves for TFAW compared to TFA is to be expected from Figure \ref{fig:TFAWdetect} as they are in the high SNR regime.

\subsection{TFAW real light curve quantitative performance}
\label{sect:tfaw_quant}

\begin{figure}[h]
   \centering
   \includegraphics[width=\linewidth,height=10.0cm,keepaspectratio]{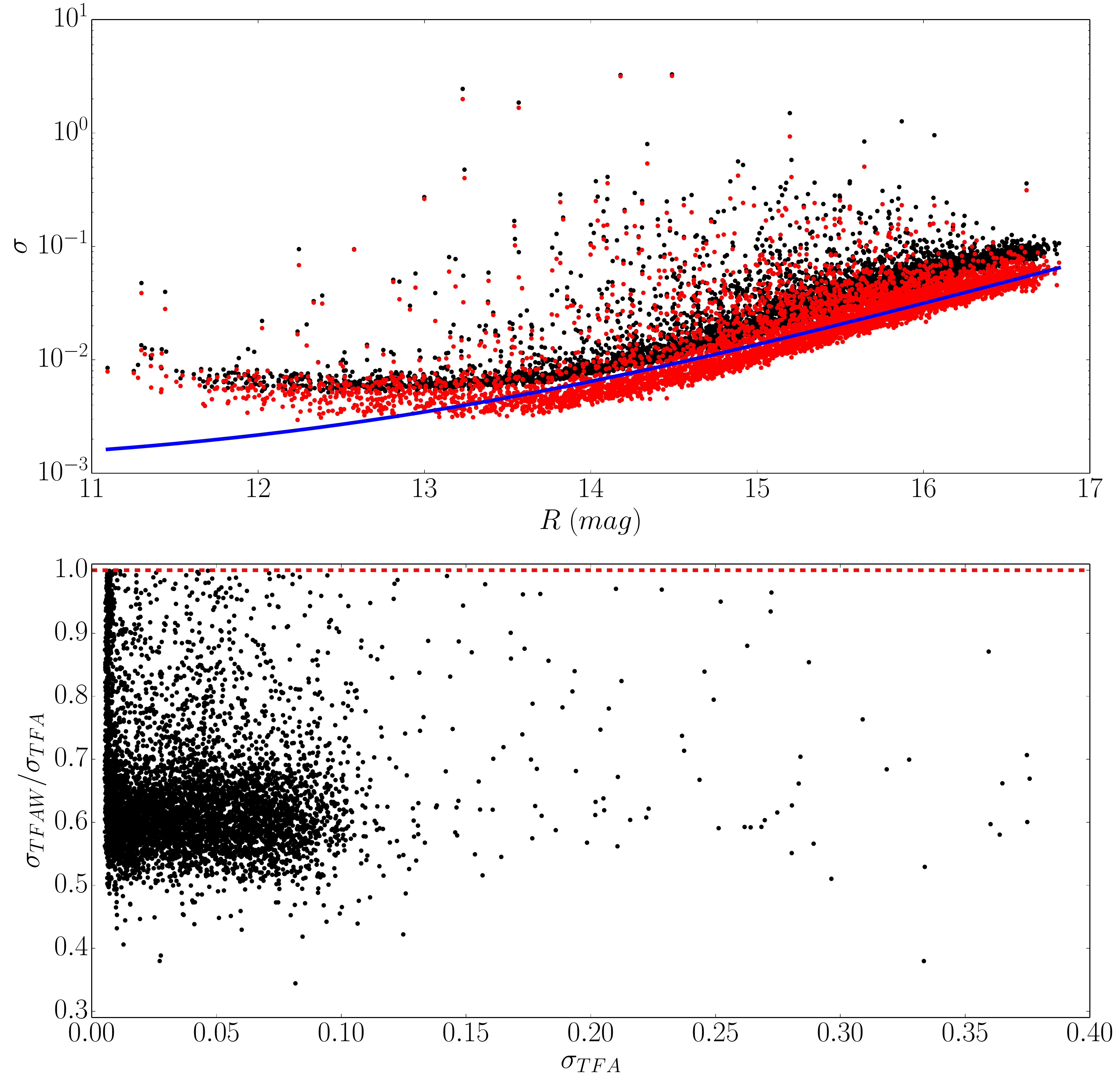}     
   \caption{\textbf{Top:} Standard deviation vs. R magnitude for TFA (black dots) and TFAW (red dots) of a set of 6485 TFRM-PSES light curves. Blue line corresponds to the sum of the scintillation, photon, background and read-out noises. \textbf{Bottom:} Decrease of the standard deviations for the same set of light curves due to the application of TFAW compared to TFA. Both TFA and TFAW results were obtained using 250 template stars. Red line corresponds to the zero correction level.}
   \label{fig:TFAWComp}
   \end{figure}

Tests made on both simulated and real light curves show that after TFAW is applied, signals affected by systematics and noise are recovered with no shape and period distortion.
Fig.~\ref{fig:TFAWComp} top panel shows the TFRM-PSES photometric precision (standard deviation $\sigma$) vs. R magnitude for both TFA and TFAW light curves. The set of the 6485 displayed light curves correspond to the same field, observed over three years. TFAW photometric performance is not only consistently better than TFA over all the R magnitude range, but also shows a closer fit with the expected blue-lined trend-free stochastic noises performance, especially the faint end.
Fig.~\ref{fig:TFAWComp} bottom panel shows, similarly to Fig.5 in~\citet{Kovacs2005}, the decrease in the standard deviations of the same sample of 6485 TFAW-filtered TFRM-PSES light curves compared to the original TFA ones. 
It can be seen, that for almost all the TFA standard deviation range (conversely brighter to fainter magnitudes), the standard deviations of TFAW light curves is $\sim$40$\%$ better than that for TFA light curves. In the case of low standard deviations (i.e. bright stars), the improvement is smaller because the noise contribution is also smaller. The presence of real variable stars in the sample could also explain a fraction of such TFA and TFAW with similar standard deviations, as their intrinsic variabilities dominate the noise contribution (see Fig.~\ref{fig:TFRM2} as an example).

The results presented in Sect.~\ref{subsect:tfrmresults} in the LS power spectra demonstrate that TFAW does not introduce false periodicities or eliminate any of the signal peaks during the frequency analysis step. For the examples presented before as well as for the rest of TFRM-PSES light curves analyzed, TFAW outlier removal and frequency analysis step results in an overall improvement of the power spectrum and of the SDEs.

Regarding the number of template stars, tests run both with simulated and real light curves show that using 250 template stars gives the best results. TFAW can also be run with less stars (<10 stars). However, the noise filtering is less efficient and some trends and systematics could still be present in the filtered data. As with the original TFA, it is not necessary to compute the template matrix for each target star, only in the cases in which the latter accidentally coincides with one of the template ones.

The algorithm has been fully implemented and parallelized in Python. The CPU time needed for a 2048 data points, single light curve TFAW analysis using 250 template stars is typically around 60 seconds and about 100 iterations. A full run of the TFAW on 6485 TFRM-PSES light curves with 2048 points per time series, 250 template stars and 90,000 BLS frequency steps takes 7.5 hours of 24~$\times$ CPUs @ 2.00 GHz. As a comparison, TFA runs the same set of light curves with the same parameters and hardware capabilities in 0.5 hours, i.e., 15$\times$ less than TFAW. The main reason for this performance loss is the extra time in the computation of the SWT and the ISWT at each iteration step. The current {\tt PyWavelets} implementation we are using for these transforms is coded in {\tt Cython}. Future effort is planned in order to significantly speed up TFAW.

\section{Conclusions}
\label{sect:conclusions}

TFAW is a totally generic, Python-based, parallelized algorithm useful for any kind of survey which seeks to improve the performance of signal detection, reconstruction and characterization, provided that a set of comparison light curves sharing the same systematics and trends as the target time series is available. TFAW makes use of an improved frequency analysis step to determine if a significant period is present in the data. This improvement consist on the introduction of a outlier-removal tool as well as the search of periods using a de-noised light curve obtained through the SWT of the original one. Once this period is found, the use of the SWT allows a more precise estimation of the signal's shape than bin averaging. The algorithm then decouples a large component of the noise contribution from the raw signal, making use of a modified version of the original TFA iterative signal reconstruction approach based on the SWT estimation of the noise at each iteration step, leading to an overall SNR improvement without alteration of the signal's time sampling or astrophysical characteristics.

Tests conducted on both simulated and real TFAW-filtered light curves show an improvement of $\sim$40$\%$ (although it can be higher) in their standard deviations with respect to the ones detrended with TFA, leading to a better characterization of the signal, without modifying its features (i.e. amplitude, phase, shape or depth). Results obtained for the frequency analysis step show that TFAW does not introduce any false periodicity, and that it can improve the overall power spectra of non-multiperiodic signals and the SDE of the peaks. In the case of simulated transits, TFAW frequency analysis step improves the transit detection rate a factor $\sim$2-5$\times$ for the low SNR signals with respect TFA and increases the SDEs up to a factor $\sim$2.5$\times$. In addition, TFAW is able to detect the transits for signals with $\sim$2$\times$ higher standard deviation. We also show that the SWT signal approximation provides a closer representation of the underlying signal with respect to bin averaging. In the case of planetary transits, this improvement is due to a better fit of the ingress and egress profiles. Also, we demonstrate that the TFAW-filtered light curve yields better MCMC posterior distributions, diminishes the bias in the fitted transit parameters and their uncertainties and narrows the credibility intervals up to a factor $\sim$10$\times$ for simulated transits.

We have also shown that TFAW is able to isolate the different underlying signals within a light curve with multiple periodic signals, such as multi-transit signals, transients, modulations or other kinds of stellar variabilities. As for single-period light curves, the SNR of each signal contribution improves due to the noise filtering capabilities of TFAW.

It is important to note that TFAW needs an even number of data points to work that will depend on the number of levels in which the signal wants to be decomposed (ideally, the length of the time series should be a power of two). Also, as described in Section~\ref{sect:tfaw}, better results are obtained if the number of SWT levels is above ten, ensuring that the noise and signal contributions are as separated as possible in the SWT decomposition of the signal. Compared to the original TFA, TFAW requires fewer reference stars to create the template used for the signal reconstruction and the noise filtering. ~\citet{Kovacs2005} suggests the use of $\sim$600 template stars, while TFAW operates stably with 250 template stars (in the data presented in this paper). 

In this paper we have applied TFAW to thousands of real light curves from TFRM-PSES survey. 

We are planning to present a dedicated TFAW test run on large numbers of The Evryscope~\citep{Law2015,Law2016a} light curves in a forthcoming paper.

TFAW algorithm also has the potential to improve the planet detection and characterization capabilities of space-based data sets, such as from CoRoT and Kepler; this study will be detailed in a forthcoming paper.

\begin{acknowledgements}
DdS acknowledges financial support from RACAB and Universitat de Barcelona. OF acknowledges funding support by the grant MDM-2014-0369 of the ICCUB (Unidad de Excelencia `Mar\'{\i}a de Maeztu'). The TFRM project is supported, in part, by the Departament d'Empresa i Coneixement de la Generalitat de Catalunya. We also acknowledge support by the Spanish Ministerio de Econom\'ia y Competitividad (MINECO/FEDER, UE) under grants AYA2013-47447-C3-1-P, AYA2016-76012-C3-1-P, MDM-2014-0369 of ICCUB (Unidad de Excelencia 'Mar\'ia de Maeztu'), and the Catalan DEC grant 2014 SGR 86. The authors would like to acknowledge the comments and suggestions of the referee that have helped to improve this work. Also, we would like to acknowledge Ignasi Ribas, Jaime Boloix, Albert Rosich, and Juan Carlos Morales for their contribution to the TFRM-PSES survey, the support team of the TFRM for the commissioning and maintenance of the telescope, as well as to José M. Solanes for contributing with his computing resources.

\end{acknowledgements}

\bibliographystyle{aa}
\bibliography{biblio}

\begin{thebibliography}{59}
\expandafter\ifx\csname natexlab\endcsname\relax\def\natexlab#1{#1}\fi

\bibitem[{{Antoja} {et~al.}(2012){Antoja}, {Helmi}, {Bienayme},
  {Bland-Hawthorn}, {Famaey}, {Freeman}, {Gibson}, {Gilmore}, {Grebel},
  {Minchev}, {Munari}, {Navarro}, {Parker}, {Reid}, {Seabroke}, {Siebert},
  {Siviero}, {Steinmetz}, {Williams}, {Wyse}, \& {Zwitter}}]{Antoja2012}
{Antoja}, T., {Helmi}, A., {Bienayme}, O., {et~al.} 2012, MNRAS, 426, L1

\bibitem[{{Arnalte-Mur} {et~al.}(2012){Arnalte-Mur}, {Labatie}, {Clerc},
  {Mart{\'{\i}}nez}, {Starck}, {Lachi{\`e}ze-Rey}, {Saar}, \&
  {Paredes}}]{Arnalte2012}
{Arnalte-Mur}, P., {Labatie}, A., {Clerc}, N., {et~al.} 2012, A\&A, 542, A34

\bibitem[{{Aschwanden} {et~al.}(1998){Aschwanden}, {Kliem}, {Schwarz},
  {Kurths}, {Dennis}, \& {Schwartz}}]{Aschwanden1998}
{Aschwanden}, M.~J., {Kliem}, B., {Schwarz}, U., {et~al.} 1998, APJ, 505, 941

\bibitem[{{Auvergne} {et~al.}(2009){Auvergne}, {Bodin}, {Boisnard}, {Buey},
  {Chaintreuil}, {Epstein}, {Jouret}, {Lam-Trong}, {Levacher}, {Magnan},
  {Perez}, {Plasson}, {Plesseria}, {Peter}, {Steller}, {Tiph{\`e}ne}, {Baglin},
  {Agogu{\'e}}, {Appourchaux}, {Barbet}, {Beaufort}, {Bellenger}, {Berlin},
  {Bernardi}, {Blouin}, {Boumier}, {Bonneau}, {Briet}, {Butler}, {Cautain},
  {Chiavassa}, {Costes}, {Cuvilho}, {Cunha-Parro}, {de Oliveira Fialho},
  {Decaudin}, {Defise}, {Djalal}, {Docclo}, {Drummond}, {Dupuis}, {Exil},
  {Faur{\'e}}, {Gaboriaud}, {Gamet}, {Gavalda}, {Grolleau}, {Gueguen},
  {Guivarc'h}, {Guterman}, {Hasiba}, {Huntzinger}, {Hustaix}, {Imbert},
  {Jeanville}, {Johlander}, {Jorda}, {Journoud}, {Karioty}, {Kerjean},
  {Lafond}, {Lapeyrere}, {Landiech}, {Larqu{\'e}}, {Laudet}, {Le Merrer},
  {Leporati}, {Leruyet}, {Levieuge}, {Llebaria}, {Martin}, {Mazy}, {Mesnager},
  {Michel}, {Moalic}, {Monjoin}, {Naudet}, {Neukirchner}, {Nguyen-Kim},
  {Ollivier}, {Orcesi}, {Ottacher}, {Oulali}, {Parisot}, {Perruchot},
  {Piacentino}, {Pinheiro da Silva}, {Platzer}, {Pontet}, {Pradines},
  {Quentin}, {Rohbeck}, {Rolland}, {Rollenhagen}, {Romagnan}, {Russ}, {Samadi},
  {Schmidt}, {Schwartz}, {Sebbag}, {Smit}, {Sunter}, {Tello}, {Toulouse},
  {Ulmer}, {Vandermarcq}, {Vergnault}, {Wallner}, {Waultier}, \&
  {Zanatta}}]{Auvergne2009}
{Auvergne}, M., {Bodin}, P., {Boisnard}, L., {et~al.} 2009, \aap, 506, 411

\bibitem[{Bakos {et~al.}(2008)Bakos, Afonso, Henning, Jordan, Holman, Noyes,
  Sackett, Sasselov, Kovács, Csubry, \& Pal}]{Bakos2009}
Bakos, G., Afonso, C., Henning, T., {et~al.} 2008, Proceedings of the
  International Astronomical Union, 4, 354

\bibitem[{Bilen \& Huzurbazar(2002)}]{Bilen2002}
Bilen, C. \& Huzurbazar, S. 2002, Journal of Computational and Graphical
  Statistics, 11, 311

\bibitem[{{Borucki} {et~al.}(2003){Borucki}, {Koch}, {Lissauer}, {Basri},
  {Caldwell}, {Cochran}, {Dunham}, {Geary}, {Latham}, {Gilliland}, {Caldwell},
  {Jenkins}, \& {Kondo}}]{Borucki2003}
{Borucki}, W.~J., {Koch}, D.~G., {Lissauer}, J.~J., {et~al.} 2003, in
  \procspie, Vol. 4854, Future EUV/UV and Visible Space Astrophysics Missions
  and Instrumentation., ed. J.~C. {Blades} \& O.~H.~W. {Siegmund}, 129--140

\bibitem[{{Bravo} {et~al.}(2014){Bravo}, {Roque}, {Estrela}, {Le{\~a}o}, \& {De
  Medeiros}}]{Bravo2014}
{Bravo}, J.~P., {Roque}, S., {Estrela}, R., {Le{\~a}o}, I.~C., \& {De
  Medeiros}, J.~R. 2014, A\&A, 568, A34

\bibitem[{{Carter} \& {Winn}(2009)}]{Carter2009}
{Carter}, J.~A. \& {Winn}, J.~N. 2009, ApJ, 704, 51

\bibitem[{Cohen {et~al.}(1992)Cohen, Daubechies, \& Feauveau}]{Cohen1992}
Cohen, A., Daubechies, I., \& Feauveau, J.-C. 1992, Communications on Pure and
  Applied Mathematics, 45, 485

\bibitem[{{Cubillos} {et~al.}(2017){Cubillos}, {Harrington}, {Loredo}, {Lust},
  {Blecic}, \& {Stemm}}]{Cubillos2017}
{Cubillos}, P., {Harrington}, J., {Loredo}, T.~J., {et~al.} 2017, AJ, 153, 3

\bibitem[{{de Freitas} {et~al.}(2010){de Freitas}, {Le{\~a}o}, {Canto Martins},
  \& {De Medeiros}}]{Freitas2010}
{de Freitas}, D.~B., {Le{\~a}o}, I.~d.~C., {Canto Martins}, B.~L., \& {De
  Medeiros}, J.~R. 2010, ArXiv e-prints [\eprint[arXiv]{1009.5090}]

\bibitem[{{del Ser} {et~al.}(2015){del Ser}, {Fors}, {N{\'u}{\~n}ez}, {Voss},
  {Rosich}, \& {Kouprianov}}]{delSer2015}
{del Ser}, D., {Fors}, O., {N{\'u}{\~n}ez}, J., {et~al.} 2015, in Astronomical
  Society of the Pacific Conference Series, Vol. 496, Living Together: Planets,
  Host Stars and Binaries, ed. S.~M. {Rucinski}, G.~{Torres}, \& M.~{Zejda},
  301

\bibitem[{Devyatkin {et~al.}(2010)Devyatkin, Gorshanov, Kouprianov, \&
  Verestchagina}]{Devyatkin2010}
Devyatkin, A.~V., Gorshanov, D.~L., Kouprianov, V.~V., \& Verestchagina, I.~A.
  2010, Solar System Research, 44, 68

\bibitem[{Donoho \& Johnstone(1994)}]{Donoho1994b}
Donoho, D.~L. \& Johnstone, I.~M. 1994, Comptes Rendus de l'Académie des
  Sciences, Série I: Mathématique, 319, 1317

\bibitem[{{Foreman-Mackey} {et~al.}(2013){Foreman-Mackey}, {Hogg}, {Lang}, \&
  {Goodman}}]{Foreman2013}
{Foreman-Mackey}, D., {Hogg}, D.~W., {Lang}, D., \& {Goodman}, J. 2013, \pasp,
  125, 306

\bibitem[{{Fors} {et~al.}(2013){Fors}, {N{\'u}{\~n}ez}, {Mui{\~n}os},
  {Montojo}, {Baena-Gall{\'e}}, {Boloix}, {Morcillo}, {Merino}, {Downey}, \&
  {Mazur}}]{Fors2013}
{Fors}, O., {N{\'u}{\~n}ez}, J., {Mui{\~n}os}, J.~L., {et~al.} 2013, PASP, 125,
  522

\bibitem[{{Fors} {et~al.}(2008){Fors}, {Richichi}, {Otazu}, \&
  {N{\'u}{\~n}ez}}]{Fors2008}
{Fors}, O., {Richichi}, A., {Otazu}, X., \& {N{\'u}{\~n}ez}, J. 2008, \aap,
  480, 297

\bibitem[{{Gim{\'e}nez de Castro} {et~al.}(2001){Gim{\'e}nez de Castro},
  {Raulin}, {Mandrini}, {Kaufmann}, \& {Magun}}]{Gimenez2001}
{Gim{\'e}nez de Castro}, C.~G., {Raulin}, J.-P., {Mandrini}, C.~H., {Kaufmann},
  P., \& {Magun}, A. 2001, A\&A, 366, 317

\bibitem[{Grané \& Veiga(2010)}]{Grane2010}
Grané, A. \& Veiga, H. 2010, Computational Statistics \& Data Analysis, 54,
  2580 , the Fifth Special Issue on Computational Econometrics

\bibitem[{{Grubbs}(1950)}]{Grubbs1950}
{Grubbs}, F.~E. 1950, Ann. Math. Statist., 21, 27

\bibitem[{{Grziwa} {et~al.}(2016){Grziwa}, {Korth}, {Paetzold}, \&
  {KEST}}]{Grziwa2016}
{Grziwa}, S., {Korth}, J., {Paetzold}, M., \& {KEST}. 2016, in AAS/Division for
  Planetary Sciences Meeting Abstracts, Vol.~48, AAS/Division for Planetary
  Sciences Meeting Abstracts, 122.02

\bibitem[{{Grziwa} {et~al.}(2014){Grziwa}, {Korth}, \&
  {P{\"a}tzold}}]{Grziwa2014}
{Grziwa}, S., {Korth}, J., \& {P{\"a}tzold}, M. 2014, European Planetary
  Science Congress 2014, EPSC Abstracts, Vol.~9, id.~EPSC2014-156, 9, EPSC2014

\bibitem[{{Grziwa} \& {P{\"a}tzold}(2016)}]{Grziwa2016a}
{Grziwa}, S. \& {P{\"a}tzold}, M. 2016, ArXiv e-prints
  [\eprint[arXiv]{1607.08417}]

\bibitem[{{Henize}(1957)}]{Henize1957}
{Henize}, K.~G. 1957, S\&T, 16

\bibitem[{Holschneider {et~al.}(1989)Holschneider, Kronland-Martinet, Morlet,
  \& Tchamitchian}]{Holschneider1989}
Holschneider, M., Kronland-Martinet, R., Morlet, J., \& Tchamitchian, P. 1989,
  A Real-Time Algorithm for Signal Analysis with the Help of the Wavelet
  Transform, ed. J.-M. Combes, A.~Grossmann, \& P.~Tchamitchian (Berlin,
  Heidelberg: Springer Berlin Heidelberg), 286--297

\bibitem[{Irwin {et~al.}(2009)Irwin, Charbonneau, Nutzman, \&
  Falco}]{irwin2009}
Irwin, J., Charbonneau, D., Nutzman, P., \& Falco, E. 2009, AIP Conference
  Proceedings, 1094, 445

\bibitem[{{Kim} {et~al.}(2009){Kim}, {Protopapas}, {Alcock}, {Byun}, \&
  {Bianco}}]{Kim2009}
{Kim}, D.-W., {Protopapas}, P., {Alcock}, C., {Byun}, Y.-I., \& {Bianco}, F.~B.
  2009, \mnras, 397, 558

\bibitem[{Knorr {et~al.}(2000)Knorr, Ng, \& Tucakov}]{Knorr2000}
Knorr, E.~M., Ng, R.~T., \& Tucakov, V. 2000, The VLDB Journal, 8, 237

\bibitem[{{Kov{\'a}cs} {et~al.}(2005){Kov{\'a}cs}, {Bakos}, \&
  {Noyes}}]{Kovacs2005}
{Kov{\'a}cs}, G., {Bakos}, G., \& {Noyes}, R.~W. 2005, MNRAS, 356, 557

\bibitem[{{Kovacs} \& {Bakos}(2008)}]{Kovacs2008}
{Kovacs}, G. \& {Bakos}, G.~A. 2008, Communications in Asteroseismology, 157,
  82

\bibitem[{{Kov{\'a}cs} {et~al.}(2002){Kov{\'a}cs}, {Zucker}, \&
  {Mazeh}}]{Kovacs2002}
{Kov{\'a}cs}, G., {Zucker}, S., \& {Mazeh}, T. 2002, A\&A, 391, 369

\bibitem[{{Kreidberg}(2015)}]{Kreidberg2015}
{Kreidberg}, L. 2015, \pasp, 127, 1161

\bibitem[{{Law} {et~al.}(2016){Law}, {Fors}, {Ratzloff}, {Corbett}, {del Ser},
  \& {Wulfken}}]{Law2016a}
{Law}, N.~M., {Fors}, O., {Ratzloff}, J., {et~al.} 2016, in Proc. SPIE, Vol.
  9906, Society of Photo-Optical Instrumentation Engineers (SPIE) Conference
  Series, 99061M

\bibitem[{{Law} {et~al.}(2015){Law}, {Fors}, {Ratzloff}, {Wulfken},
  {Kavanaugh}, {Sitar}, {Pruett}, {Birchard}, {Barlow}, {Cannon}, {Cenko},
  {Dunlap}, {Kraus}, \& {Maccarone}}]{Law2015}
{Law}, N.~M., {Fors}, O., {Ratzloff}, J., {et~al.} 2015, PASP, 127, 234

\bibitem[{{Machado} {et~al.}(2013){Machado}, {Leonard}, {Starck}, {Abdalla}, \&
  {Jouvel}}]{Machado2013}
{Machado}, D.~P., {Leonard}, A., {Starck}, J.-L., {Abdalla}, F.~B., \&
  {Jouvel}, S. 2013, A\&A, 560, A83

\bibitem[{Mallat(2008)}]{Mallat2008}
Mallat, S. 2008, A Wavelet Tour of Signal Processing: The Sparse Way (Elsevier
  Science)

\bibitem[{Mallat \& Hwang(1992)}]{Mallat1992}
Mallat, S. \& Hwang, W.~L. 1992, IEEE transactions on information theory, 38,
  617

\bibitem[{Mallat(1989)}]{Mallat1989}
Mallat, S.~G. 1989, IEEE Transactions on Pattern Analysis and Machine
  Intelligence, 11, 674

\bibitem[{{Mandel} \& {Agol}(2002)}]{Mandel2002}
{Mandel}, K. \& {Agol}, E. 2002, ApJ, 580, L171

\bibitem[{Meyers {et~al.}(1993)Meyers, Kelly, \& O'Brien}]{Meyers1993}
Meyers, S.~D., Kelly, S.~D., \& O'Brien, J.~J. 1993, Monthly Weather Review,
  121, 2858

\bibitem[{{Moudden} {et~al.}(2005){Moudden}, {Cardoso}, {Starck}, \&
  {Delabrouille}}]{Moudden2005}
{Moudden}, Y., {Cardoso}, J.-F., {Starck}, J.-L., \& {Delabrouille}, J. 2005,
  EURASIP Journal on Applied Signal Processing, 2005, 484606

\bibitem[{{N{\'u}{\~n}ez} \& {Otazu}(1996)}]{Nunez1996}
{N{\'u}{\~n}ez}, J. \& {Otazu}, X. 1996, Vistas in Astronomy, 40, 555

\bibitem[{{Otazu} {et~al.}(2002){Otazu}, {Rib{\'o}}, {Peracaula}, {Paredes}, \&
  {N{\'u}{\~n}ez}}]{Otazu2002}
{Otazu}, X., {Rib{\'o}}, M., {Peracaula}, M., {Paredes}, J.~M., \&
  {N{\'u}{\~n}ez}, J. 2002, MNRAS, 333, 365

\bibitem[{{Peirce}(1852)}]{Peirce1852}
{Peirce}, B. 1852, AJ, 2, 161

\bibitem[{{Petigura} \& {Marcy}(2012)}]{Petigura2012}
{Petigura}, E.~A. \& {Marcy}, G.~W. 2012, \pasp, 124, 1073

\bibitem[{{Pollacco} {et~al.}(2006){Pollacco}, {Skillen}, {Collier Cameron},
  {Christian}, {Hellier}, {Irwin}, {Lister}, {Street}, {West}, {Anderson},
  {Clarkson}, {Deeg}, {Enoch}, {Evans}, {Fitzsimmons}, {Haswell}, {Hodgkin},
  {Horne}, {Kane}, {Keenan}, {Maxted}, {Norton}, {Osborne}, {Parley}, {Ryans},
  {Smalley}, {Wheatley}, \& {Wilson}}]{Pollacco2006}
{Pollacco}, D.~L., {Skillen}, I., {Collier Cameron}, A., {et~al.} 2006, PASP,
  118, 1407

\bibitem[{{R{\'e}gulo} {et~al.}(2007){R{\'e}gulo}, {Almenara}, {Alonso},
  {Deeg}, \& {Roca Cort{\'e}s}}]{Regulo2007}
{R{\'e}gulo}, C., {Almenara}, J.~M., {Alonso}, R., {Deeg}, H., \& {Roca
  Cort{\'e}s}, T. 2007, \aap, 467, 1345

\bibitem[{{Scargle}(1982)}]{Scargle1982}
{Scargle}, J.~D. 1982, \apj, 263, 835

\bibitem[{Starck {et~al.}(1998)Starck, Murtagh, \& Bijaoui}]{Starck1998}
Starck, J., Murtagh, F., \& Bijaoui, A. 1998, Image Processing and Data
  Analysis: The Multiscale Approach (Cambridge University Press)

\bibitem[{{Starck} \& {Murtagh}(1994)}]{Starck1994}
{Starck}, J.-L. \& {Murtagh}, F. 1994, A\&A, 288, 342

\bibitem[{{Szatmary} {et~al.}(1994){Szatmary}, {Vinko}, \&
  {Gal}}]{Szatmary1994}
{Szatmary}, K., {Vinko}, J., \& {Gal}, J. 1994, A\&AS, 108

\bibitem[{{Tamuz} {et~al.}(2005){Tamuz}, {Mazeh}, \& {Zucker}}]{Tamuz2005}
{Tamuz}, O., {Mazeh}, T., \& {Zucker}, S. 2005, MNRAS, 356, 1466

\bibitem[{Torrence \& Compo(1998)}]{Torrence1998}
Torrence, C. \& Compo, G.~P. 1998, Bulletin of the American Meteorological
  society, 79, 61

\bibitem[{Tukey(1949)}]{Tukey1949}
Tukey, J.~W. 1949, Biometrics, 5, 99

\bibitem[{{Voss}(2006)}]{Voss2006}
{Voss}, H. 2006, PhD thesis, Technischen Universität Berlin

\bibitem[{Waldmann(2014)}]{Waldmann2014}
Waldmann, I.~P. 2014, The Astrophysical Journal, 780, 23

\bibitem[{{Wheatley} {et~al.}(2013){Wheatley}, {Pollacco}, {Queloz}, {Rauer},
  {Watson}, {West}, {Chazelas}, {Louden}, {Walker}, {Bannister}, {Bento},
  {Burleigh}, {Cabrera}, {Eigm{\"u}ller}, {Erikson}, {Genolet}, {Goad},
  {Grange}, {Jord{\'a}n}, {Lawrie}, {McCormac}, \& {Neveu}}]{Wheatley2013}
{Wheatley}, P.~J., {Pollacco}, D.~L., {Queloz}, D., {et~al.} 2013, in European
  Physical Journal Web of Conferences, Vol.~47, European Physical Journal Web
  of Conferences, 13002

\bibitem[{{Wo{\'z}niak} {et~al.}(2004){Wo{\'z}niak}, {Vestrand}, {Akerlof},
  {Balsano}, {Bloch}, {Casperson}, {Fletcher}, {Gisler}, {Kehoe}, {Kinemuchi},
  {Lee}, {Marshall}, {McGowan}, {McKay}, {Rykoff}, {Smith}, {Szymanski}, \&
  {Wren}}]{Wozniak2004}
{Wo{\'z}niak}, P.~R., {Vestrand}, W.~T., {Akerlof}, C.~W., {et~al.} 2004, \aj,
  127, 2436

\end{thebibliography}

\end{document}